\documentclass[]{aa}  
\usepackage{graphicx}
\usepackage{txfonts}
\usepackage{xcolor}
\usepackage{amsmath}
\usepackage{multirow}
\usepackage{lipsum}

\usepackage[colorlinks=true,
            linkcolor=blue,
            citecolor=blue,
            filecolor=black,
            urlcolor=blue]{hyperref}

\title{A multiwavelength study of an early galaxy group merger in COSMOS revealed by two tailed radio galaxies at z = 0.35}
\titlerunning{Multiwavelength study of an early galaxy group merger in COSMOS with two tailed radio galaxies}

\author{P. Vuli\'c
          \inst{1}\fnmsep\thanks{Email: pvulic@phy.hr}
          \and
          V. Smol\v{c}i\'c\inst{1}\fnmsep\thanks{Email: vs@phy.hr}
          \and
          G. Gozaliasl\inst{2,4}
          \and
          I. Delvecchio\inst{3}
          \and
          A. Finoguenov \inst{4}
          }

\institute{Department of Physics, University of Zagreb, Bijeni\v{c}ka cesta 32, Zagreb, Croatia
        \and
            Department of Computer Science, Aalto University, PO Box 15400, 00076, Espoo, Finland
        \and
            INAF – Osservatorio di Astrofisica e Scienza dello Spazio di Bologna, Via Gobetti 93/3, I-40129 Bologna, Italy
        \and
            Department of Physics, University of Helsinki, P.O. Box 64, FI-00014 Helsinki, Finland
            }

\date{Received July 7 2025 / Accepted September 23 2025}
 
\abstract{We report the discovery of two tailed radio galaxies in the COSMOS field, associated with a massive, dynamically unrelaxed galaxy group detected in X-rays at z = 0.349. One of them is a wide-angle tail (WAT) galaxy, supporting the role of WATs as tracers of dynamically young groups and clusters. Our multiwavelength analysis combines VLA radio data, HST-ACS imaging, COSMOS2020 photometric redshifts, COSMOS2015 photometry, the newest compilation of spectroscopic redshifts in COSMOS, and X-ray observations from Chandra and XMM-Newton. We used these data to study the tailed radio galaxies, their host galaxies, and the group environment. Both radio galaxies are hosted by massive ($\log_{10}(M_*/M_{\odot})=11.88\pm0.03$ and $\log_{10}(M_*/M_{\odot})=11.49\pm0.06$), red, elliptical galaxies with extended stellar halos, as revealed by a color, magnitude, and stellar mass analysis combined with GALFIT modeling and surface-brightness profiles. One corresponds to the brightest group galaxy (BGG), while the other is the second-brightest. A diffuse intragroup medium (IGM) is characterized by its irregular shape and the analysis of the X-ray spectra of the group core reveals high temperature ($T_X=2.4\pm0.6\hspace{0.1cm}\mathrm{keV}$) and an electron density of $(8.2\pm0.3)\times 10^{-4}\hspace{0.1cm}\mathrm{cm^{-3}}$. A galaxy overdensity associated with the group was detected via Voronoi tessellation, using COSMOS2020 CLASSIC photometric redshifts, displaying an irregular morphology, along with evidence of substructure. Assuming the jet bending results from interaction with the IGM, we find a high relative velocity between the BGG and the IGM ($v_{\mathrm{BGG/IGM}} \gtrsim 540$ km/s), primarily due to bulk gas motion. Our findings indicate a dynamically young system in the early stages of assembly via group-group merging.}
\keywords{Galaxies: clusters: intracluster medium - Galaxies: groups: general - Galaxies: jets - Cosmology: observations - X-rays: galaxies: clusters - Radio continuum: galaxies}
\begin{document}
\maketitle
\section{Introduction}
\label{sec:intro}
Galaxies tend to form gravitationally bound structures such as galaxy groups or clusters, which present the building blocks of the large-scale structure in the Universe. The dynamics of groups and clusters and mergers between them are therefore directly connected to the growth of cosmic structures, in agreement with the cold dark matter (CDM) cosmological model.
Clusters and groups found in a virialized (dynamically relaxed) state have a regularly shaped, spherically symmetric, galaxy distribution exhibiting a central concentration. Non-relaxed galaxy groups and clusters, which are likely a result of recent merger processes, are often characterized by irregular distribution of galaxies and of the intragroup or intracluster medium (IGM or ICM). Non-regular systems are ideal for studying galaxy dynamics and interaction and can provide insight into gravitational interactions on large scales.\par
\citet{best2005} showed that radio loud active galactic nuclei (AGNs) prefer to reside in dense environments commonly known as groups and clusters. Radio galaxies are often found to be hosted by the brightest cluster or group galaxy (BCG or BGG), which is typically a massive elliptical galaxy \citep{best2007, croft2007}. In regular clusters or groups, the BCG or BGG is usually found at the bottom of the gravitational potential well dominating the system's dynamics. Hence, small peculiar velocities relative to the cluster or group mean are expected and observed for these galaxies in relaxed systems; for instance, a line-of-sight (LoS) $\Delta v\lesssim 150\hspace{0.1cm}\mathrm{km/s}$ for relaxed, poor clusters or massive groups \citep{beers1995, Ow2001}. Values consistent with the above were also found in the more recent study of BCG peculiar velocities in the local universe (for regular systems, if extended to the group regime) by \citet{lauer2014}. However, the situation is different in unrelaxed systems, where larger peculiar velocities are observed and are related to the existence of galaxy substructures \citep{bird94}. Wide-angle tail (WAT) galaxies are a class of radio galaxies that have their jets bent at a large angle forming a C shape, while in narrow-angle tail (NAT) galaxies, jets are bent much more sharply, creating a U shape \citep{miley1972,RO1976}. Both WATs and NATs are commonly found in galaxy groups and clusters and have previously been suggested as indicators of dynamic systems \citep{smolcic2007, oklopcic2010, douglass2011, dasadia2016}. The most plausible explanation for the observed radio jet bending is that it emerges from their interaction with the ICM or IGM \citep{begelman1979}. This requires a high velocity of their host galaxy with respect to the medium, which can be achieved in non-regular systems or, alternatively, a high medium density would be required.\par
\begin{figure*}
    \centering
    \includegraphics[width=0.6\hsize]{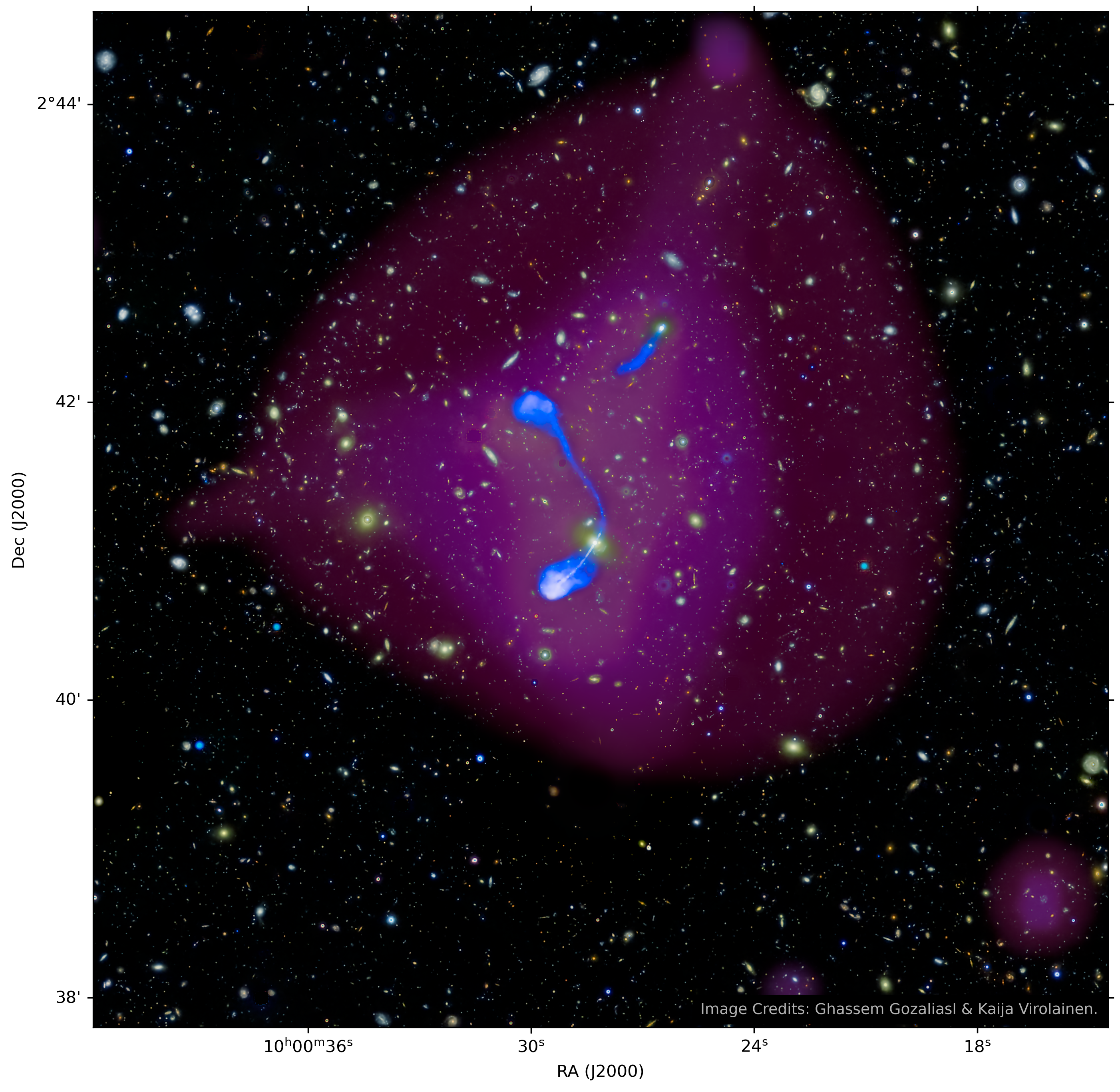}
    \caption{Multiwavelength view of our massive galaxy group in the COSMOS field. The background is an RGB composite created from Subaru \textit{i}-, \textit{r}-, and \textit{g}-band observations \citep{capak2008}, revealing the optical galaxy distribution. The magenta overlay indicates extended X-ray emission from the intragroup medium, derived from wavelet-filtered 0.5–2 keV data from Chandra and XMM–Newton \citep[see][]{Gozaliasl}, tracing diffuse hot gas on scales of 16–256 arcseconds. The blue structures highlight the radio emission from the radio galaxies 10913 (WAT, the brightest group galaxy) and 44 (tailed, the second brightest group galaxy), tracing radio jets launched from their central engines. The emission is based on the 3 GHz map from the VLA-COSMOS 3 GHz Large Project \citep{3ghz_smol_a}. Together, the image captures the interaction between galaxies, the radio AGNs, and the surrounding hot plasma in a galaxy group at a redshift of $z = 0.349$.}
    \label{fig:combined}
\end{figure*}

\vspace{-0.03cm}
The system analyzed here is a galaxy group at z=0.349, which was previously detected in the \textit{Cosmic Evolution Survey} (COSMOS, \citealt{cosmos})\footnote{\url{https://cosmos.astro.caltech.edu/page/astronomers}} field in X-ray \citep{Gozaliasl}. A total of seven radio sources have been found within this galaxy group; two of them, hereafter labeled 10913 and 44 based on their 3 GHz catalog ID \citep{3ghz_smol_a}, correspond to tailed radio galaxies (see below for details). The other five, labeled 4092, 1549, 10366, 633, and 527,  are point-like sources (see Table~\ref{table:srces} for details). In Fig.~\ref{fig:combined}. we show a multiwavelength RGB - X-ray - radio composite image of the system (see caption for details).\par
In this paper (hereafter, Paper I) we present a multiwavelength analysis of the system, focusing on the environment. In Section \ref{sec:data}, we introduce the data, in Section \ref{sec:xray_analysis}, we focus on X-ray properties of the system, and in Section \ref{sec:optical_properties_env}, we describe our optical and near-infrared (NIR) analysis of the environment. In Section \ref{sec:radio_src_and_hosts}, we analyze the radio galaxies and their hosts. A discussion is presented in Section \ref{sec:discussion}, along with a short summary in Section \ref{sec:summary}.
Throughout this paper, we use $H_0=69.32\hspace{0.15cm}\mathrm{km\hspace{0.1cm}Mpc^{-1}\hspace{0.1cm}s^{-1}}$, $\Omega_M=0.2865$,
and $\Omega_{\lambda}=0.7135$. We define the radio spectral index as positive, assuming $F_\nu\propto \nu^{-\alpha}$.

\vspace{-0.1cm}
\section{Data}
\label{sec:data}
To analyze the system, we used data from the COSMOS \citep{cosmos} field at different wavelengths, including radio, optical/NIR, and X-ray data.
\vspace{-0.2cm}
\subsection{Radio data}
\label{sec:radiodata}
We used the VLA-COSMOS 3 GHz Large Project data \citep{3ghz_smol_a}. The 3 GHz continuum mosaic (hereafter, the 3 GHz map) was constructed from the total of 384 h of VLA observations toward the entire $2\hspace{0.1cm}\mathrm{deg}^2$ COSMOS field, yielding a $0.75^{\prime\prime}$ resolution, and an average of $2.3\hspace{0.1cm}\mathrm{\mu Jy/beam}$ in terms of the root mean square ($rms$) noise. We also used the associated $rms$ noise map, radio source catalog, and multiwavelength counterpart catalog (\citealt{3ghz_smol_b}, hereafter VLA 3 GHz catalog).

\vspace{-0.1cm}
\subsection{Optical/NIR imaging data}
\label{sec:opticalnirdata}
We used imaging data observed through the F814W broad-band filter with the \textit{Wide Field Channel} (WFC) detector of the \textit{Advanced Camera for Surveys} (ACS) mounted on the Hubble Space Telescope (HST).
A large portion of the COSMOS field ($1.6\hspace{0.1cm}\mathrm{deg}^{2}$) was covered with 583 orbits of HST
observations in two observing cycles from 2003 to 2005 \citep{koekemoer2007}. We used both unrotated tiles and the final rotated mosaic (north up, \citealt{koekemoer2007,massey2010}).
We also used Subaru \textit{i}-, \textit{r}-, and \textit{g}-band imaging data \citep{capak2008}.

\vspace{-0.1cm}
\subsection{Optical/NIR photometry and photometric redshifts}
\label{sec:photometrydata}
We used photometric redshifts from the COSMOS2020 CLASSIC catalog (\citealt{COSMOS2020}, hereafter CLASSIC).
The source detection and multiwavelength photometry were performed including the imaging data collected in the COSMOS field since the last public catalog \citep{Laigle}. It gathers multiwavelength photometry data from different surveys, ranging from the  NIR ($\approx150\hspace{0.1cm}\mathrm{nm}$) to near-ultraviolet (NUV, $\approx8\hspace{0.1cm}\mu\mathrm{m}$) (for more details, see Table I in \citealp{COSMOS2020}). We used photometric redshifts derived by \citealp{COSMOS2020} using the above-described data, employing LePhare code \citep{ilbert2006}. The CLASSIC LePhare photometric redshift precision $\sigma$ is $0.008$ for sources with Hyper Supreme-Cam (HSC) $i$-band magnitudes $i<22.5$ and better than $0.015$, $0.024$, and $0.044$ for $i<24$, $i<25$, and $i<27$ , respectively (see Figure 13. in \citealt{COSMOS2020}). The value of $\sigma$ was calculated from the differences between the photometric and the spectroscopic redshifts, $\Delta z$, of sources within a given $i$ magnitude range, according to \cite{COSMOS2020}:
 \begin{equation}
     \sigma=1.48\times\mathrm{median}\bigg(\frac{\lvert\Delta z - \mathrm{median}(\Delta z)\rvert}{(1+z_{spec})}\bigg)
 .\end{equation}
 Apart from the photometric redshifts from CLASSIC, we also used the less numerous, but more reliable spectroscopic redshifts from the most recent COSMOS spectroscopic redshift catalog \citep{Khostovan2025}. This catalog represents a public spectroscopic COSMOS archive, gathering known spectroscopic redshifts in the COSMOS field, originating from different surveys. We used only secure spectroscopic redshifts (quality flag 3 or 4) that were also flagged as PUBLIC.\par
 We further used optical photometry data from the COSMOS2015 catalog \citep[][hereafter COSMOS2015]{Laigle}, namely,\ Subaru Supreme-Cam (SC): $B$, $r$, and $i^{+}$ broad band magnitudes originating from measurements presented by \cite{Taniguchi}, given that the photometry for our brightest source (the main host galaxy) is not available in the CLASSIC catalog.
 The source detection algorithm used in CLASSIC performs poorly in the vicinity of masked regions and our brightest source of interest lies on the edge of one such region. For more details about masked regions and the source detection algorithm used in CLASSIC, we refer to \cite{COSMOS2020}. The photometry of COSMOS2015 is highly consistent with the one from CLASSIC (see Appendix C in \citealt{COSMOS2020}).
 
\vspace{-0.1cm}
\subsection{X-ray data}
\label{sec:xraydata}
We used the combined Chandra and XMM-Newton 0.5–2 keV data of our group and the X-ray group catalog of \citet{Gozaliasl} (also see references therein). These authors used all the available X-ray observations performed by the Chandra \citep{weisskopf2000} and XMM-Newton \citep{jansen2001} observatories to search for galaxy clusters and groups in the COSMOS $2$~deg$^2$ field. In total, they identified  247 X-ray groups down to an X-ray flux limit of $3\times10^{-16}$~erg~cm$^{-2}$~s$^{-1}$. The mass and redshift ranges of the groups and clusters are $M_{200c}=8\times10^{12}-3\times10^{14}\hspace{0.1cm}M_{\odot}$, and $z=0.08-1.53$, respectively (for more details see  \citet{Gozaliasl} and references therein). We also use 0.5-7 keV EPIC-pn data from the XMM-Newton wide field survey of the COSMOS field \citep{hasinger2007}.
Furthermore, we used the Chandra COSMOS-Legacy Survey catalog of X-ray point sources \citep{civano2016}. They detected 4016 X-ray point-like sources based on the data from a combined 4.6 Ms Chandra program on the $2.2\hspace{0.1cm}\mathrm{deg}^2$ of the COSMOS field. We also used the corresponding catalog of optical and infrared (IR) counterparts \citep{marchesi2016}, containing a counterpart and photometric redshift for $\approx 97\%$ of 4016 X-ray sources. The latter are found using LePhare \citep{ilbert2006}, assuming specific priors and template libraries, chosen on the basis  of the X-ray flux and other morphological and photometric parameters of the detected sources. \citet{marchesi2016} also provided spectroscopic redshifts for 54\% of them. We found no significant increase in the number of publicly available, reliable spectroscopic redshifts when cross-matching with the new catalog of spectroscopic redshifts in the COSMOS \citep{Khostovan2025}.
\section{X-ray properties}
\label{sec:xray_analysis}
\subsection{The galaxy group}
\cite{Gozaliasl} reported an X-ray detection of our galaxy group of interest.
We adopted the basic group properties determined by \cite{Gozaliasl}, listed in Table~\ref{table:xray_group_properties} in Appendix \ref{app:tables}. This group was also one of 10 COSMOS galaxy groups studied by \citet{Kettula}. Their work included both X-ray and weak lensing analysis to calibrate the scaling relation between X-ray spectroscopic temperature and weak lensing mass in the group mass regime. Group and cluster masses are commonly found from thermal X-ray emission under the assumption of hydrostatic equilibrium (scaling relations), however, gravitational lensing enables a direct measurement of their masses regardless of dynamical state. Given our group is suspected to be a non-relaxed system, we adopted the weak lensing mass estimated by \citet{Kettula} and listed their estimates of $M_{200}$, $M_{500}$, and $T_X$ (from X-ray spectra) in Table~\ref{table:xray_group_properties}. Values from the two different studies presented in Table~\ref{table:xray_group_properties} have been re-scaled to the cosmology we use in this paper (if needed). The two sets of ($T_X$, $M_{200}$) values are consistent within the reported uncertainties. The group seems to be very luminous, massive, and hot, namely, $L_X$ is $\approx 750\%$, $M_{200}$ is $\approx 320\%$, and $T_X$ is $\approx 130\%$ above the median values for the control sample, consisting of X-ray detected galaxy groups in the COSMOS field in the redshift bin [0.1,0.5] (using the data of \citealt{Gozaliasl}).

\vspace{-0.4cm}
\subsection{Galaxy group:\ Analysis of the IGM}
\label{sec:galaxygroup_analysis_of_the_IGM}
To analyze the properties of the IGM, we use 0.5–7.0 keV energy band of XMM-Newton EPIC-pn data from a single pointing (OBSID 0203361101, \citealt{hasinger2007}). The spectrum was extracted for the group core, using a circle of radius $1.225'$ (corresponding to $\sim$367 kpc), encompassing the location of both radio galaxies: 10913 and 44. The spectral fitting was performed using the absorbed APEC model in XSPEC, with the Galactic hydrogen column density fixed to $N_\mathrm{H} = 1.8 \times 10^{20}$ cm$^{-2}$ and the metal abundance fixed to 0.2 solar relative to \citet{AG1989}. Errors on fitted parameters are quoted at the 68\% confidence level ($1\sigma$). The best-fit temperature is $T_X=2.4\pm 0.6$ keV. Assuming $N_\mathrm{e} = 1.2 N_\mathrm{p}$, we estimated the electron density to be $n_\mathrm{e} = (8.2 \pm 0.3) \times 10^{-4}$ cm$^{-3}$ from the normalization of APEC. $\chi^2$ statistics is used for minimization with at least 30 counts per bin in source or background spectrum, which was taken from the same observation. The fit yielded a reduced chi-square $\chi_\mathrm{r}^2$ of $\sim1.0$, indicating a statistically acceptable result and the results are consistent with the values reported by \cite{Kettula}, who used a slightly different extraction region to study the scaling relations. The extended emission is morphologically irregular: elongated in the north-south direction in the group core, aligned with the orientation of the jets of the WAT galaxy 10913, and transitioning to an east–west elongation beyond the core. The results of this section are further discussed in the context of the group's dynamics in Section \ref{sec:unrelaxed_group}.

\vspace{-0.2cm}
\subsection{X-ray point sources within the galaxy group}
\label{sec:xray_ps_within_the_group} 
We checked the catalog of optical/IR counterparts of Chandra point sources (\citealt{marchesi2016}; see Section \ref{sec:xraydata}) for all X-ray point sources within the detected galaxy group.
We confined our search to a circular area in the plane of the sky with radius $R_{200}$ (Table~\ref{table:xray_group_properties}), and we only kept the sources with photometric redshift in the range $z \in [\overline{z}-3\sigma(1+\overline{z}), \overline{z}+3\sigma(1+\overline{z})]$, where $\sigma=0.03$ is the corresponding photometric redshift error (\citealt{marchesi2016}). The central value of $\overline{z}=0.349$ is the galaxy group's redshift (see Table~\ref{table:xray_group_properties} here, and \citealt{Gozaliasl}).\par
We found five X-ray point sources within our group, listed in Table~\ref{table:xray_sources} in Appendix \ref{app:tables} and sorted with their X-ray flux (flux\_f) in descending order. 
The first source corresponds to tailed galaxy 44, and the third to WAT galaxy 10913, within $3^{\prime\prime}$ search radius. The second-brightest X-ray source, lid\_183, matches radio source 528 (ID from 3 GHz catalog, see Table \ref{table:srces}), initially not included among the radio sources within the group according to the search criteria presented in Section \ref{sec:radio_src_and_hosts}. However, as described there in more details, this source most probably also belongs to our system of interest. The other two have no counterpart in radio sources within the group.

\vspace{-0.2cm}
\section{Optical properties of the environment}
\label{sec:optical_properties_env}
\subsection{Voronoi tessellation analysis}
\label{sec:voronoi}
\subsubsection{Method}
We here analyze the spatial distribution of optical/NIR sources from the CLASSIC catalog (\citealt{COSMOS2020}, Section \ref{sec:opticalnirdata}) within the $10^{\prime}\times10^{\prime}$ neighborhood of the radio sources in the plane of the sky. The sources extracted and used in the analysis are mostly classified as galaxies in the CLASSIC catalog ($\mathrm{lp\_type=0}$) and a small portion of them are classified as X-ray sources ($\mathrm{lp\_type=2}$). Given that in CLASSIC there are no reliable photometric redshifts  for X-ray sources, the latter are only included in the analysis if there is available photometric redshift from the catalog of optical/IR counterparts of Chandra point sources (\citealt{marchesi2016}; see Section \ref{sec:xraydata}). We set the center of this $10^{\prime}\times10^{\prime}$ area to correspond to the center of the galaxy group detected in X-ray (see Table~\ref{table:xray_group_properties} and \citealt{Gozaliasl}). This group is at the approximately same redshift as the radio sources ($z\approx 0.35$). We limited our analysis to galaxies in non-masked areas (FLAG\_COMBINED=0, see \citealt{COSMOS2020}) with an HSC $i$ band magnitude lower than $24$. It is worth noting that no significant qualitative difference appears in the final results if we include fainter sources ($i<25$). We only consider sources with photometric redshifts in the range $z \in [\overline{z}-3\sigma(1+\overline{z}), \overline{z}+3\sigma(1+\overline{z})]$, where for galaxies $\sigma=0.015$ is the corresponding photometric redshift error derived taking into account the magnitude limit set above (see Section \ref{sec:opticalnirdata} and/or \citealt{COSMOS2020} for details), and for X-ray sources $\sigma=0.03$ is the photometric redshift error from \citet{marchesi2016}. We took $\overline{z}=0.349$, corresponding to the group redshift \citep{Gozaliasl}. We emphasize that, due to the large photometric redshift uncertainty (compared to the precision of coordinates in the plane of the sky) the physical spatial dimension of the above redshift range is much larger than the one corresponding to the extent of the sample in the plane of the sky. This broad redshift range is necessary to ensure a good level of completeness for the potential group members sample.

We performed a 2D Voronoi tessellation (VT) on the above presented $10^{\prime}\times10^{\prime}$ galaxy sample, using the \texttt{Voronoi} module from \texttt{scipy.spatial} subpackage in Python. VT is a non-parametric method which divides the (Ra, Dec) plane into smaller cells, each assigned to a single galaxy, that contain all spatial points closer to that particular galaxy than to any other galaxy in the plane. By construction, cell areas are lower in the regions of increased galaxy density. Each cell can be assigned with a local source density value that corresponds to the inverse of the cell's area,
\begin{equation}
    \rho_{i}=\frac{1}{A_{i}}
,\end{equation}
where $i \in [1,N]$, and N is a total number of cells, namely, galaxies in the sample. Due to its non-parametric nature, this method is not only sensitive to symmetric, but it can also detect elongated and irregular structures, as demonstrated by \citet{Eb}, who developed a VT-based method to detect X-ray sources as overdensities of photons following the Poisson distribution. \citet{Ramella} used this method to detect galaxy clusters as overdensities of galaxies following the same distribution in space. Here, this method enables us to quantitatively describe and compare local source densities in different parts of the plane, and therefore, to possibly identify one or more over-dense regions that may correspond to the galaxy group (previously detected in X-ray) and (or) its substructures.

To avoid boundary effects, we only worked with cells which have all vertices within $10^{\prime}\times 10^{\prime}$ region of interest.
To extract the cells with increased source density in the $10^{\prime}\times10^{\prime}$ environment, we first determined the density threshold $\rho_{th}$, above which we consider the galaxies as located in overdense regions ($\rho_{i}>\rho_{th}$). 
\begin{figure}[]
    \centering
    \includegraphics[width=0.85\hsize]{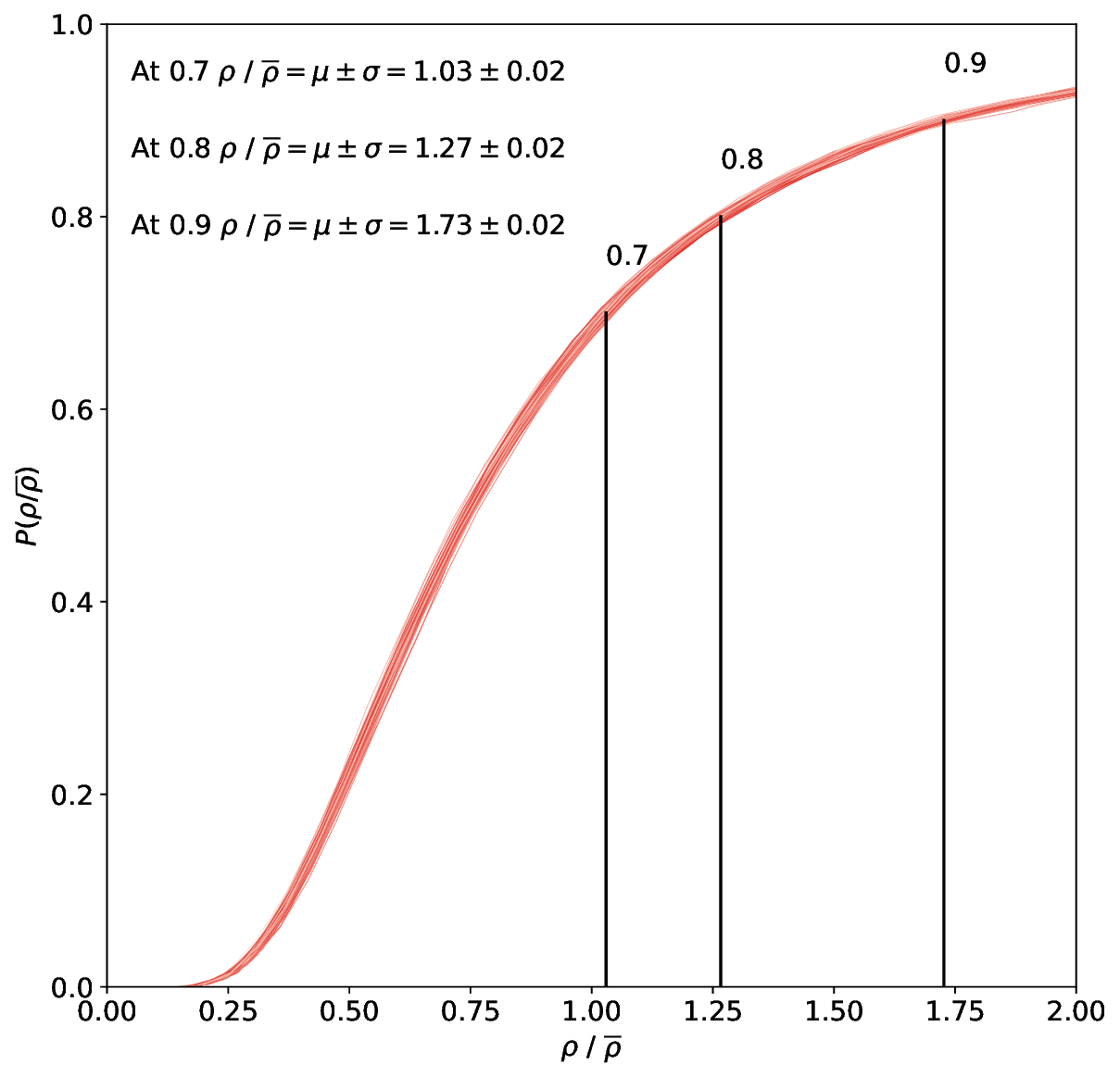}
    \caption{Cumulative distribution functions of the source (galaxy) density scaled with the mean value of source density (for a given sample) for 100 simulated background samples (red curves). Vertical black lines represent the mean density thresholds at $P=0.7$, $P=0.8$, and $P=0.9$ based on 100 simulations.}
    \label{fig:CDFs_sim}%
\end{figure}
To find this threshold, we started by assuming that the background galaxies are distributed randomly in space. We performed 100 simulations, generating a large sample of galaxies in each
one, namely, randomly generating galaxy Ra and Dec coordinates while assuming a uniform probability density function. The number of these sources per unit of non-masked area in the simulation is set to the value calculated for the entire COSMOS field within the above given redshift and magnitude range. For each of these 100 samples, we performed VT, calculated cell source densities and created the corresponding cumulative distribution function (CDF), expressed as
\begin{equation}
    F\hspace{0.1cm}(\rho/\hspace{0.05cm}\overline{\rho})=P\hspace{0.1cm}(\rho^{\prime}/\hspace{0.05cm}\overline{\rho}\le \rho/\hspace{0.05cm}\overline{\rho})
,\end{equation}
where $\overline{\rho}$ is the mean value of source density for a given sample of galaxies. These are shown in Fig.~\ref{fig:CDFs_sim}.
From the CDFs, we found the density threshold that we used in the further analysis as $\rho/\hspace{0.05cm}\overline{\rho}$, for which $P\hspace{0.1cm}(\rho^{\prime}/\hspace{0.05cm}\overline{\rho} \le \rho/\hspace{0.05cm}\overline{\rho})=0.8$ (i.e., which is larger than $80\%$ of the density values from the sample). We also determined the density thresholds larger than $70\%$ and $90\%$ of density values from the sample. The mean threshold values originating from 100 simulations and the corresponding standard deviations for $P=0.7$, $P=0.8$, and $P=0.9$ are: $\rho/\hspace{0.05cm}\overline{\rho}= 1.03 \pm 0.02$, $\rho/\hspace{0.05cm}\overline{\rho} = 1.27 \pm 0.02$, and $\rho/\hspace{0.05cm}\overline{\rho} = 1.73 \pm 0.02$, respectively.
Finally, we filtered the Voronoi cells by their local source density, taking into account the thresholds determined.

\subsubsection{Results}
\label{sec:voronoi_results}
\begin{figure}
    \centering
    \includegraphics[width=\hsize]{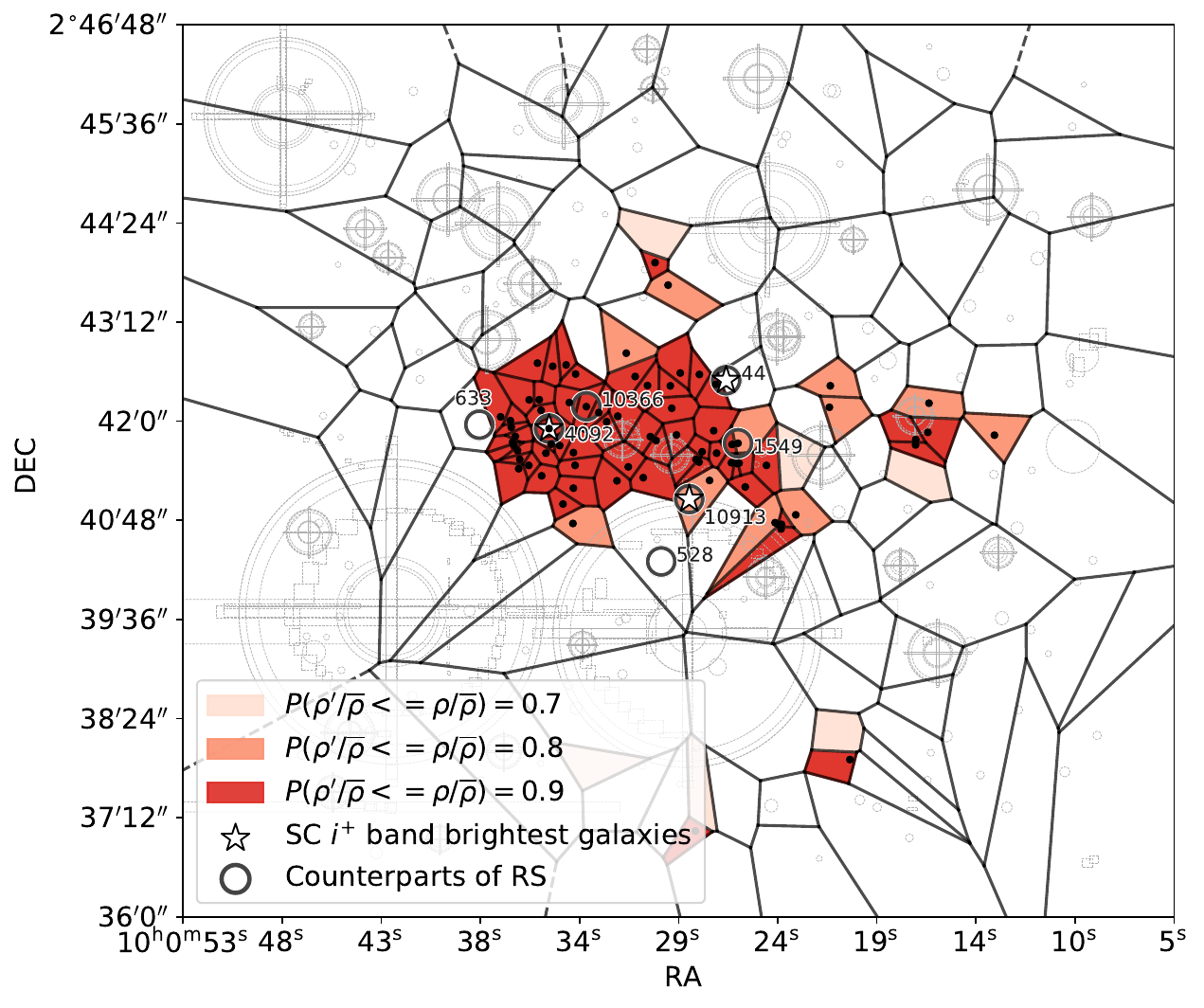}
    \caption{Results of our Voronoi tesselation analysis:\ Colored cells are left after filtering according to the pre-calculated source density thresholds (0.7, 0.8, and 0.9 in different colors; see Section \ref{sec:voronoi} for details). Black dots are galaxies with the local source density higher than $80\%$ of the density values from the total sample, and together with the hosts of radio galaxies 10913 and 44 (manually added, see the text for details) form a sample of probable group members. Host galaxies of the 7 radio sources are shown as open gray circles.  10913, 44, and 4092 correspond to the first, second, and third brightest galaxies of the group, respectively, in the SC $i^{+}$ band (white stars).}
    \label{fig:voronoi2D}
\end{figure}

The results of the VT analysis, described in detail in the previous section, are shown in Fig.~\ref{fig:voronoi2D}. The cells passing through different of the three filters are represented by different colors. As expected from the existence of a galaxy group previously detected in X-ray, a central overdensity is noticeable in the resulting image. It seems to be mostly formed by the galaxies, that is, cells with the local source density higher than $90\%$. We note that the optical counterpart of radio galaxy 10913 is missing from all three samples. This is due to the fact that it is  not present in the CLASSIC catalog (before any filtering; see Section \ref{sec:opticalnirdata} for details). On the contrary, the counterpart source of the radio galaxy 44 is present in CLASSIC and it is not located within a masked area; however, its Voronoi cell does not pass any of the above density criteria since this source is on the edge of the overdense structure detected in Fig.~\ref{fig:voronoi2D}. Therefore,  we find it is associated with lower local source density. Nevertheless, it is reasonable to expect these two sources to belong to the detected central overdensity since radio galaxies are regularly found hosted by galaxy groups or clusters. For this reason, we added by hand these two sources to the sample of group candidates at 80\% which makes the total count of 76 sources (74 from the results of VT and 2 added by hand). This is the sample of probable group members which we use in the further analysis (for creating color-magnitude and color-stellar mass diagrams). The described sample also includes the optical counterparts of radio sources 4092, 1549, and 10366, which fulfill the highest (4092 and 10366) and the second highest (1549) density criteria set above. The counterpart of radio source 633 is found near the eastern edge of the overdensity and does not pass any of the density criteria. This, combined with its spectroscopic redshift (see Table~\ref{table:srces}), suggests 633 may belong to a different galaxy structure at lower redshifts. The counterpart of 528 is in masked area and therefore it was not possible to determine the local galaxy density around this source. In Fig.~\ref{fig:voronoi2D}, the counterparts of the radio sources are marked with dark-gray circles and we highlight (white stars) the three brightest sources in SC $i^{+}$ band based on the analysis of Section \ref{sec:cmd_analysis}. We note that all three of them correspond to the optical counterparts of radio sources, from the first to the third brightest: 10913 (WAT), 44 (tailed), and 4092. Inspecting Fig.~\ref{fig:voronoi2D}, we  notice that the hosts of both radio galaxies, 10913 (WAT, BGG) and 44 (tailed, the second brightest), are located at the periphery of the group-related overdensity, which is consistent with the scenario of a non-virialized galaxy group. However, we note that 10913 lies at the very edge of a masked region.\par
We find weak signs of substructuring within the detected overdensity, with a few arguably dense accumulations around radio sources 4092, 10366, and 1549. An exception is an extremely dense, elongated accumulation of galaxies in the far eastern part of the group-related overdensity. It is possible that this accumulation is related with another galaxy structure (group) around radio source 633, also found far east in the VT diagram; however, at lower spectroscopic redshift $(z\sim0.3,$ as already discussed above). We also detected a small structure in the south, detached from the large group-related overdensity and consisting of only two galaxies. These are unlikely to be the group members, given the lack of the observed IGM X-ray emission in this region (see Fig.~\ref{fig:combined}). It is also possible they are still infalling into the group's gravitational potential. However, if they belonged to the group, this would imply there may be other members hidden behind the large masked region in the south, right below the detected group overdensity. We note that, in general, the conclusions in this section may be influenced by the possible presence of additional member galaxies obscured by the masked regions (shown in light-gray color in Fig.~\ref{fig:voronoi2D}).
\vspace{-0.1cm}
\subsection{Color-magnitude diagram}
\label{sec:cmd_analysis}
We created a color-apparent magnitude diagram (CMD) to examine the properties of stellar population in 76 galaxies considered as probable group members (as defined in Section \ref{sec:voronoi_results}).
\begin{figure}[]
    \centering
    \includegraphics[width=0.48\textwidth]{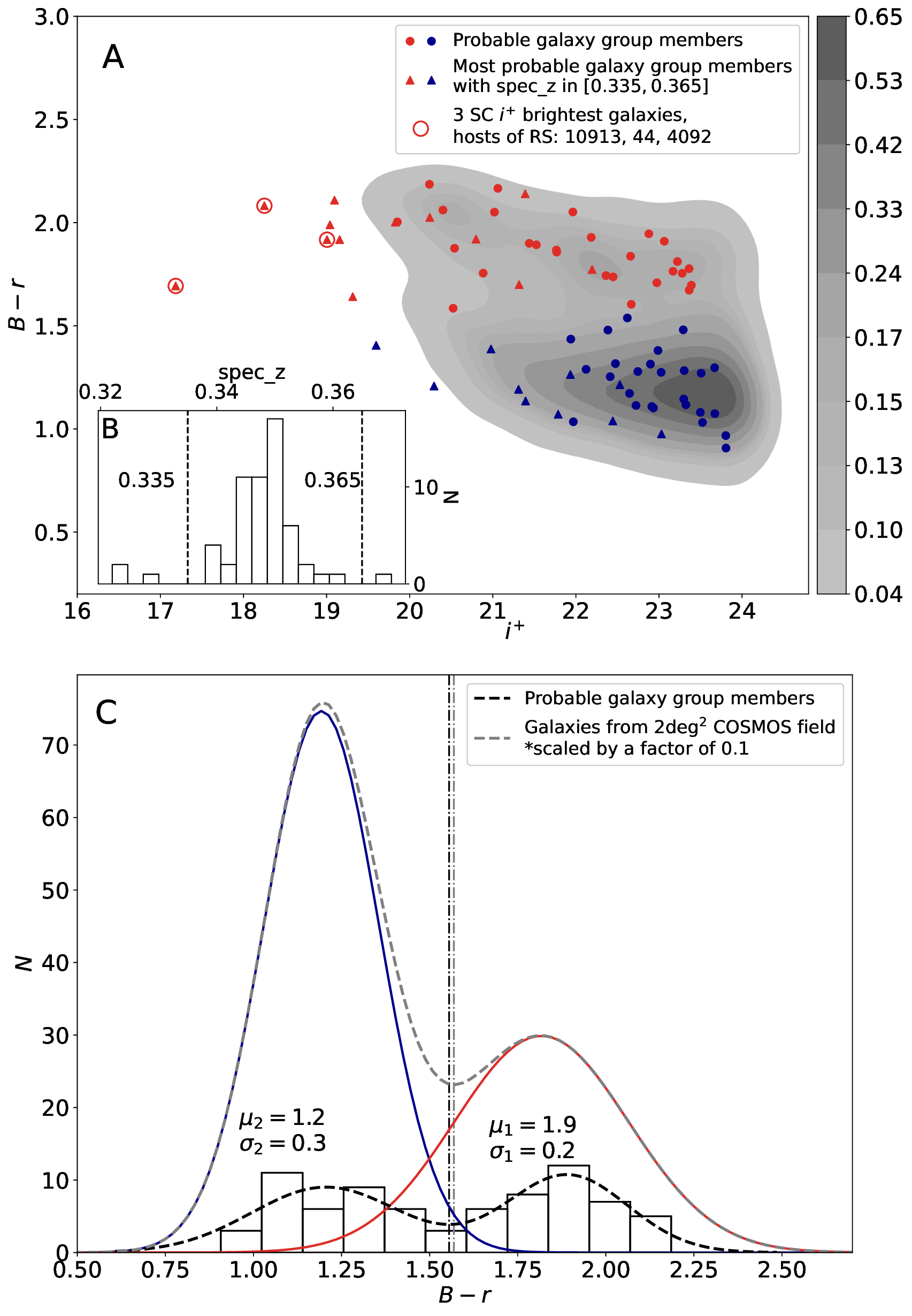}
    \caption{Panel A: CMD for 76 probable group members. Triangles represent the 23 most probable members with high-confidence spectroscopic redshifts in a narrow range [0.335, 0.365] centered on the group's redshift (see panel B, shown as an inset of panel A). Galaxies are separated in blue and red according to color-determination scheme explained in Section \ref{sec:defining_the_red_sequence}. Panel C: 76 probable group galaxies' bimodal distribution in color compared to the one of the large COSMOS-0.35 sample. Dot-dashed vertical lines indicate the inflection points of the two distributions.}
    \label{fig:CMD_B-r}
\end{figure}
\vspace{-0.3cm}
\subsubsection{Galaxy sample and photometry}
\label{sec:cmd_galaxysampleandphotometry}
The CMD sample consists of 76 probable group members (see  Section \ref{sec:voronoi_results}). We use Subaru SC $B$, $r$, and $i^{+}$ broad band apparent magnitudes (\citealt{Taniguchi}; also see Section \ref{sec:photometrydata} for details).
We chose the apparent (rather than the absolute) magnitudes to avoid introducing additional uncertainties.
The $B-r$ vs.\  $i^{+}$ CMD is shown in panel A of Fig.~\ref{fig:CMD_B-r}.
The contours shown in the background visualize the probability density function in 2D color-magnitude plane for the entire $2\hspace{0.1cm}\mathrm{deg}^2$ COSMOS field, in the same redshift bin considered here (COSMOS-0.35 sample hereafter).
Different levels here correspond to iso-proportions of the density; for instance, the contour at 0.3 means that 30\% of the distribution lies below it (in the area outside of it).
In addition, by cross-matching the sample of probable group members with the catalog of spectroscopic redshifts \citep{Khostovan2025}, we find that 26 (out of 76, $\approx 35\%$) sources have a high-confidence spectroscopic counterpart ($Q_{f}=3$ or $Q_{f}=4$).
Among these 26 sources, 23 fall within a narrow range ($0.335<\mathrm{spec\_z}<0.365$) around the group's X-ray center $\overline{z}=0.349$ and, thus, they are considered the most probable group members. The width of this range is set to match the one of the distribution of high quality ($Q_{f}=3$ or $Q_{f}=4$) spectroscopic redshifts from the $10^{\prime}\times 10^{\prime}$ region of interest forming a (group-related) peak at $z\approx 0.35$ (see panel B, shown as an inset of panel A in Fig.~\ref{fig:CMD_B-r}). The 23 most probable group members and other probable members are marked in the CMD with triangles and circles, respectively (panel A of Fig.~\ref{fig:CMD_B-r}).
\vspace{-0.2cm}
\subsubsection{Identifying the red sequence}
\label{sec:defining_the_red_sequence}
\vspace{-0.1cm}
Probable group members are bimodally distributed in the color-magnitude space, as it is generally expected for galaxies \citep{strateva2001}. A large share of red galaxies, however, is expected among them, considering that elliptical (early type) galaxies predominantly reside in denser regions of the universe; namely, there is a large fraction of early type galaxies in groups and clusters \citep{Dressler}. Moreover, these galaxies are expected to form a long, nearly horizontal structure in the CMD, namely, the red sequence \citep{Baum,VS}. To investigate this, we separate the old, red, bright galaxies from younger, blue, and less bright galaxies. As a reference for comparison we used the COSMOS-0.35 sample described above. By inspecting the bimodal distribution  of the latter in color, we find that it can be modeled with the sum of two Gaussian functions (blue and red lines and gray dashed line for their sum in panel C of Fig.~\ref{fig:colorsm}). The same can be applied to the bimodal distribution in color of the probable group members sample consisting of 76 sources (black dashed line, panel C of Fig.~\ref{fig:colorsm}). We find no galaxies between the two inflection points $\mathrm{lim}_{\mathrm{low}}$, $\mathrm{lim}_{\mathrm{high}}$ (vertical black dot-dashed lines in panel C of Fig.~\ref{fig:colorsm}), corresponding to the two distributions, and thus we select red and blue galaxies based on the $B-r>\mathrm{lim}_{\mathrm{high}}$ and $B-r<\mathrm{lim}_{\mathrm{low}}$ criteria, respectively. As expected, by comparing the color histograms and the corresponding Gaussian sum functions (panel C in Fig.~\ref{fig:colorsm}), the probable group members sample has a larger fraction of red galaxies compared to the COSMOS-0.35 sample. These red galaxies form a structure resembling a red sequence. The three brightest (from the first to the third) sources in the CMD are the optical counterparts of radio sources 10913 (WAT), 44 (tailed) and 4092, respectively.

\vspace{-0.2cm}
\subsection{Color-stellar mass diagram}
\begin{figure}[]
    \centering
    \includegraphics[width=0.87\hsize]{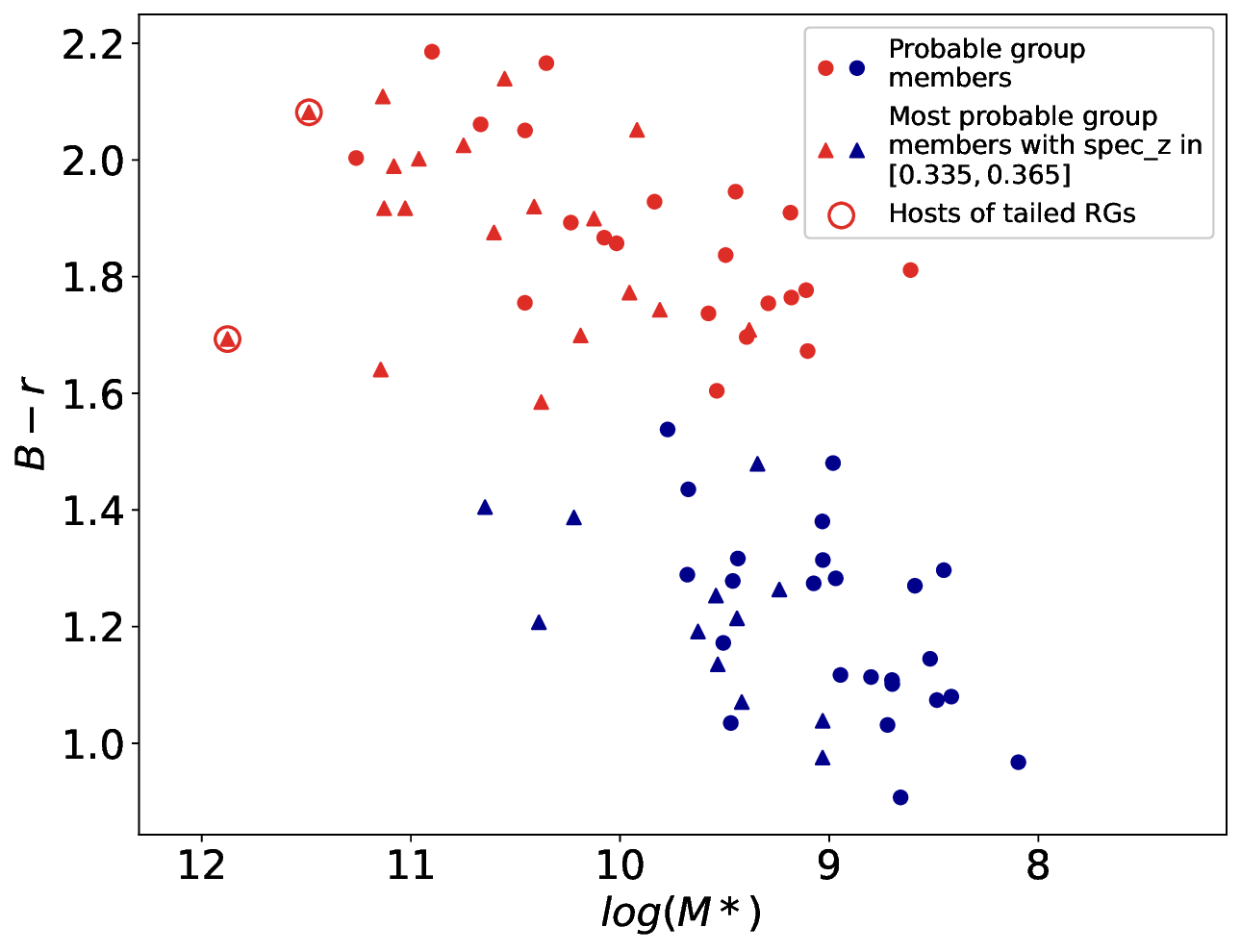}
    \caption{CSMD for 76 probable group members. Stellar masses are calculated as explained in Section \ref{sec:csmd_analysis}. The first and the second most massive galaxies correspond to the hosts of radio galaxies 10913 (WAT) and 44 (tailed), respectively. Markers and colors follow the scheme described in the caption of Fig.~\ref{fig:CMD_B-r}.}
    \label{fig:colorsm}
\end{figure}

\vspace{-0.1cm}
\label{sec:csmd_analysis}
We create a color-stellar mass diagram (CSMD) for the sample of 76 galaxies that are probable group members. Here, we used $B-r$  color indices (photometry of \citealt{Taniguchi}) and stellar masses calculated by fitting a spectral energy distribution (SED) of each galaxy in the sample. We used the MAGPHYS SED-fitting tool \citep{dacunha2008}, which is designed to reproduce a variety of galaxy SEDs, from weakly star-forming to starbursting galaxies. It relies on the energy balance between the dust-absorbed stellar continuum (from \citealt{BC2003}), and the reprocessed dust emission at IR wavelengths (from \citealt{CF2000}). Stellar mass estimates are given in \citet{chabrier2003} initial mass function (IMF) and their typical 1$\sigma$ uncertainty is of the order of 0.1~dex. The CSMD is shown in Fig.~\ref{fig:colorsm}. The two most massive group galaxies correspond to the two brightest in the same order, namely, hosts of WAT radio galaxy 10913 ($\log_{10}(M_*/M_{\odot})=11.88\pm0.03$) and tailed radio galaxy 44 ($\log_{10}(M_*/M_{\odot})=11.49\pm0.06$).

\section{Radio sources and their host galaxies}
\label{sec:radio_src_and_hosts}
\renewcommand{\arraystretch}{1.2}
\begin{table}
\caption{Flux densities and monochromatic luminosities at 3 GHz for radio sources: 10913 (WAT, BBG), 44 , 4092, 1549, 10366, 633, and 528.}
\label{table:fluxlum}      
\centering
\scalebox{0.85}{
\begin{tabular}{c | c | c |c}    
\hline\hline
 Source & $F_{\mathrm{3GHz}}\hspace{0.2cm}[\mathrm{mJy}]$ & $\alpha$\tablefootmark{$\star$} & $L_{\mathrm{3GHz}}\hspace{0.2cm}[10^{24}\hspace{0.1cm}\mathrm{W\hspace{0.05cm}Hz^{-1}}]$ \\ \hline                  
   10913 & $30\substack{+4\\-2}$ & $0.78\pm0.10$& $11.9\substack{+1.6 \\ -0.8}$\\
   44 & $2.8\substack{+1.0 \\ -0.4}$ & $0.56\pm0.16$ & $1.0\substack{+0.4 \\ -0.1}$\\
   4092 & $0.028\pm0.003$ & $0.73\pm0.35$ & $0.011\pm0.002$\\
   1549 &  $0.071\pm0.004$ & $0.73\pm0.35$ & $0.028\pm0.003$\\ 
   10366 & $0.011\pm0.002$ & $0.73\pm0.35$ & $0.005\pm0.001$\\
   633 & $0.121\pm0.007$ & $0.73\pm0.35$ & $0.035\pm0.004$\\
   528 & $0.152\pm0.008$ & $0.73\pm0.35$ & $0.061\pm0.007$ \\ \hline
\end{tabular}
}
\tablefoot{
\tablefoottext{$\star$}{Spectral index as calculated in Paper II (Vuli\'c et al., in prep.) based on radio flux densities at 5 different frequencies from 325 MHz to 3 GHz (for radio galaxies 10913 and 44) and assumed $0.73\pm0.35$ for point sources (see Section \ref{sec:radiomorph_and_lum} for details), corresponding to the average value found for the full 3 GHz population \citep{3ghz_smol_a}.}
}
\end{table}
\renewcommand{\arraystretch}{1}

\subsection{Radio sources within the group}
We found five radio sources within the group by cross-matching (within $1^{\prime\prime}$ radius) our CLASSIC VT input sample of galaxies (see Section \ref{sec:voronoi}), confined to $R_{200}$ (see Table~\ref{table:xray_group_properties}) with the VLA 3 GHz catalog \citep{3ghz_smol_b}. This method is limited to finding only the radio sources within the group which have a non-masked optical/NIR counterpart. As previously explained in Section \ref{sec:photometrydata}, our WAT radio galaxy 10913 has no counterpart in the CLASSIC; however, it is reasonable to expect it within the group. Moreover, the systematic search for X-ray point sources within the group (Section \ref{sec:xray_ps_within_the_group}) reveals a source within $R_{200}$, with a spectroscopic redshift $z= 0.345$, also active in radio, but without CLASSIC optical/NIR counterpart due to the same reasons as for radio galaxy 10913 - contamination by a bright foreground star. Therefore, we add these two manually to the sample of radio sources within the group, which yields a total of seven (see Table~\ref{table:srces} in Appendix \ref{app:tables}).
Two of these seven are tailed radio galaxies (10913 and 44) and the remainder point-like radio sources.
\vspace{-0.2cm}
\subsection{Radio morphology and luminosities}
\vspace{-0.1cm}
\label{sec:radiomorph_and_lum}
\begin{figure*}[]
    \centering
    \includegraphics[width=0.85\textwidth]{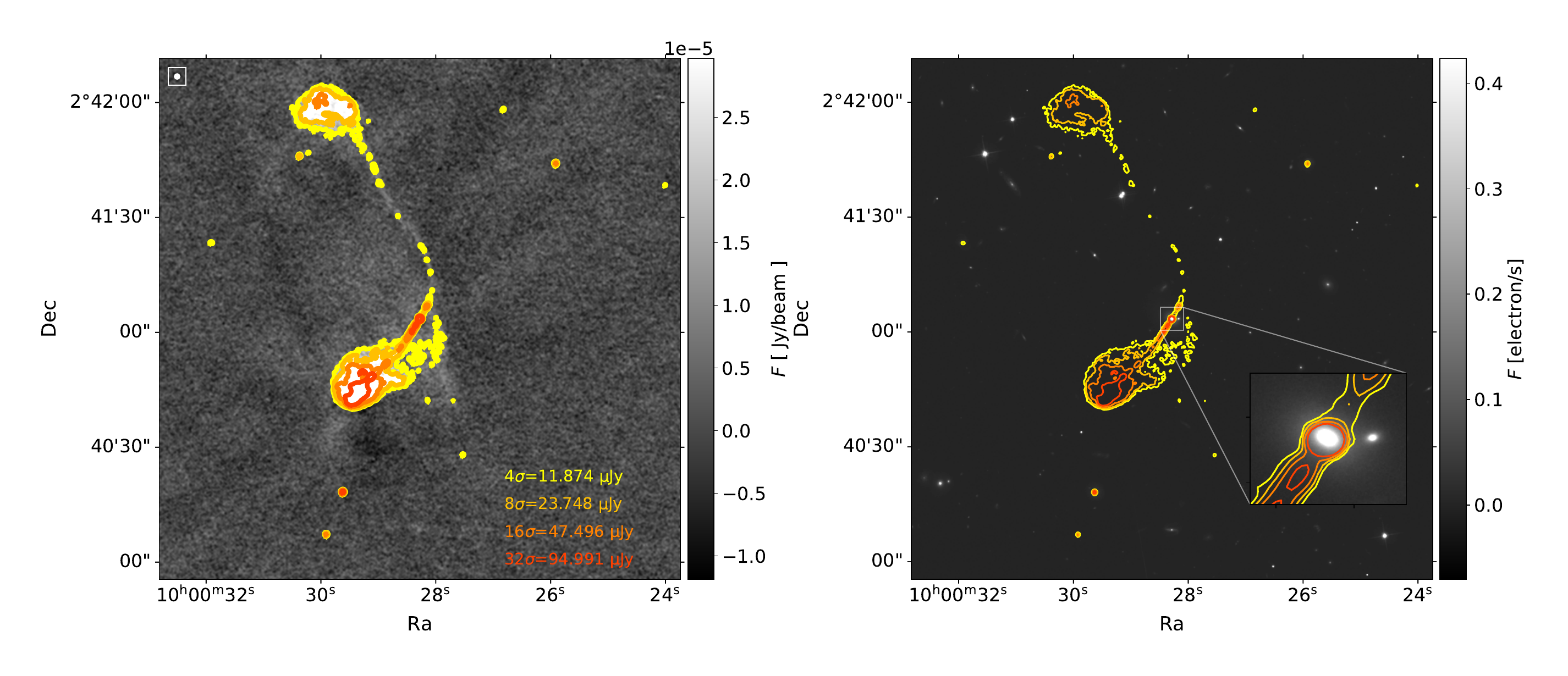}
    \caption{Left: Radio image of WAT radio galaxy 10913 - cutout from 3 GHz map \citep{3ghz_smol_a}. Resolution is $0.75^{\prime\prime}$, and the average local $rms$ noise $\sigma=2.97\hspace{0.1cm}\mathrm{\mu Jy/beam}$. Radio contours are shown and the details on contour levels are given in the bottom right corner while in the upper left corner the corresponding radio beam is shown. Contour levels are set as $n\sigma$, while $n$ goes over integer values in logarithmic ($\mathrm{log_{2}}$) scale. The map scale is from $-4\hspace{0.05cm}\sigma$ to $10\hspace{0.05cm}\sigma$. Right: 3 GHz radio contours overlaid on the HST ACS F814W image (host galaxy in optical, \citealt{koekemoer2007}).}
    \label{fig:10913_3ghz_HST}
\end{figure*}
\begin{figure*}[]
    \centering
    \includegraphics[width=0.85\textwidth]{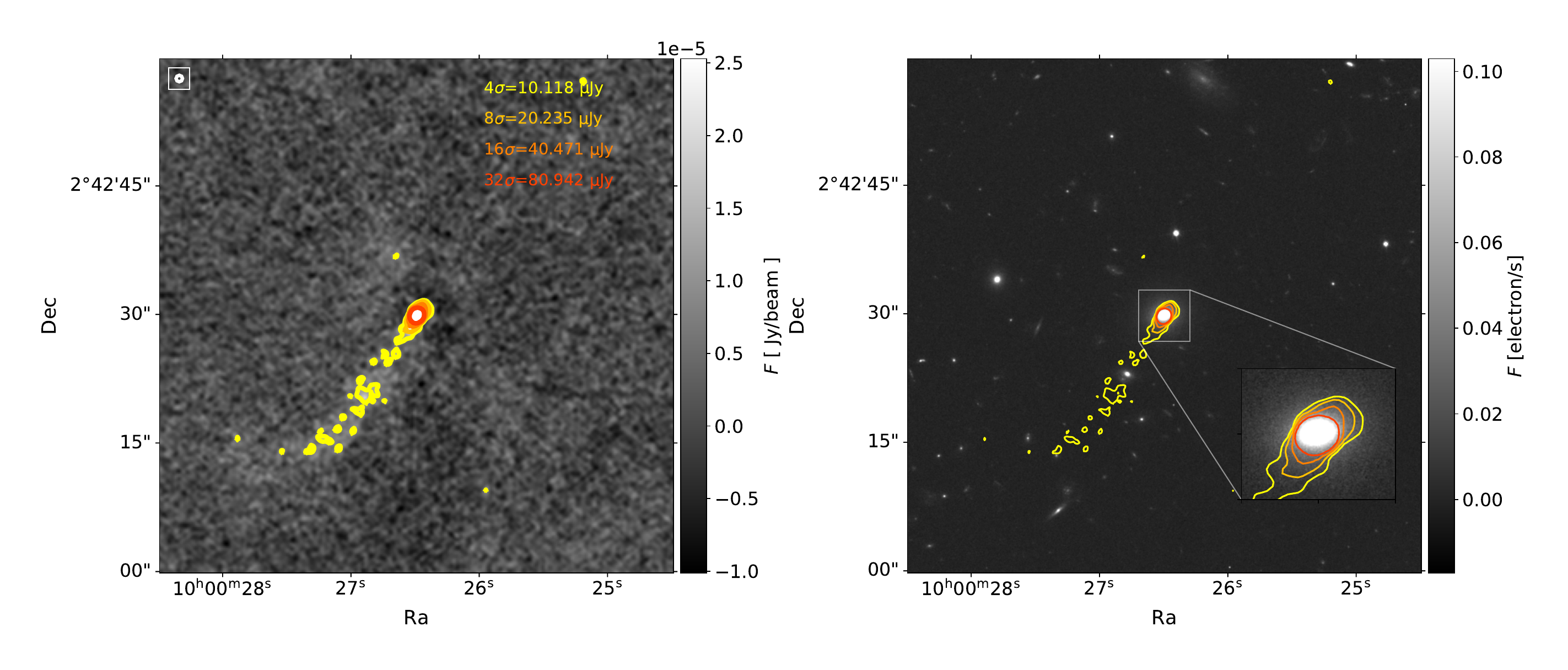}
    \caption{Left: Radio image of tailed radio galaxy 44 - cutout from 3 GHz map \citep{3ghz_smol_a}, shown with radio contours following $\mathrm{log_{2}}$ scale (see description of Fig.~\ref{fig:10913_3ghz_HST} for details on contour levels). Resolution is $0.75^{\prime\prime}$, and the average local $rms$ noise $\sigma=2.53\hspace{0.1cm}\mathrm{\mu Jy/beam}$. The map scale is from $-4\hspace{0.05cm}\sigma$ to $10\hspace{0.05cm}\sigma$. Right: 3 GHz radio contours overlaid on the HST ACS F814W image (host galaxy in optical, \citealt{koekemoer2007}).}
    \label{fig:44_3ghz_HST}
\end{figure*}
In Figs.~\ref{fig:10913_3ghz_HST} and \ref{fig:44_3ghz_HST}, we present cutouts from the 3 GHz radio map \citep{3ghz_smol_a}, with the corresponding radio contours (left) and the contours overlaid over the HST F814W mosaic cutout (\citealt{koekemoer2007}, right) for the two tailed radio galaxies, 10913 and 44, respectively. Contour levels follow logarithmic scale in terms of $\sigma$, the average local $rms$ noise calculated by averaging values at different pixels in the corresponding $rms$ noise map cutout.
Radio contours indicate 10913 is an FRII source \citep{FR1974}, and 44 is an FRI. Radio galaxy 44 shows only one detected jet (tail), suggesting that it is most probably a head-tail (HT) radio galaxy, possibly a NAT galaxy, where resolution ($\sim 4\hspace{0.1cm}\mathrm{kpc}$ at the group's redshift) is insufficient to resolve the two bent jets. This and other scenarios are further discussed in Section \ref{sec:headtail_or_wat_44}. Throughout the paper, we refer to it simply as a tailed radio galaxy.
We calculate radio luminosities $L_{\mathrm{3GHz}}$ at 3 GHz for all seven radio sources using the following expression,
\begin{equation}
L_{\mathrm{3GHz}}=\frac{4\pi F_{\mathrm{3GHz}} D_L^2}{(1+z)^{1-\alpha}}
.\end{equation}
Here, $z$ is the spectroscopic redshift (Table~\ref{table:srces}), $D_L$ is the luminosity distance calculated using $z$, and $\alpha$ is the radio spectral index. For radio galaxies 10913 and 44, we derived $\alpha$ from radio data at five frequencies (from 325 MHz to 3 GHz) as a part of a detailed radio analysis presented in Paper II (Vuli\'c et al., in prep.). For point sources, we adopted $\alpha = 0.73 \pm 0.35$, the average spectral index for the full 3 GHz population \citep{3ghz_smol_a}, since they were detected with low signal-to-noise ratios (S/N$\lesssim3$) at all frequencies available to us (except at 3 GHz), preventing a reliable individual spectral index calculation. The 3 GHz flux density $F_{\mathrm{3GHz}}$, and its associated error were adopted from the VLA 3 GHz catalog \citep{3ghz_smol_a} for point sources. For the two radio galaxies, we recalculated the flux densities in a consistent way and estimated their errors. We note that for the 3 GHz catalog, BLOBCAT was used to derive the flux densities of single-component sources such as 44, while flux densities of multi-component sources, such as 10913, were derived manually without errors. We found $F_{\mathrm{3GHz}}$ using the 3 GHz continuum map \citep{3ghz_smol_a}, integrating over pixels with flux densities above $3\sigma$, and dividing the sum with the beam size in pixels. Here, $\sigma$ is the local $rms$ noise for a given source. To estimate uncertainties, we repeat this at $2\sigma$ and $4\sigma$ thresholds, and use deviations from the $3\sigma$ value as the upper and lower flux density errors, respectively. Although this method may overestimate the errors, we use it to account not only for calibration uncertainties, but also for potential ambiguities arising from the multiple flux density threshold options that must be selected in the analysis. Our measurements are consistent with catalog values for 10913 and marginally consistent for 44 (for comparison see Fig.~\ref{Fig:compare_flux} in Appendix \ref{app:compare_flux}). The luminosity errors were calculated by propagating errors of flux density and spectral index. The resulting flux densities and luminosities are listed in Table~\ref{table:fluxlum}.
\vspace{-0.25cm}
\subsection{GALFIT analysis of light profiles of host galaxies}
\label{sec:galfit_short}
To describe the light profiles of tailed radio galaxies' hosts, we fit 2D models to the F814W HST-WFC-ACS optical/NIR imaging data \citep{koekemoer2007}, using  GALFIT \citep{peng2010}. We also performed this analysis for the host galaxy of radio source 4092, motivated by its high brightness at these wavelengths (see Table~\ref{table:srces}, as well as the results of Section \ref{sec:cmd_analysis}) and a high local galaxy density in its vicinity (see the results of VT, Section \ref{sec:voronoi_results}). Here, we outline this analysis and present the main results. For full details, see appendix \ref{app:Galfit}.

We first extracted and prepared galaxy cutouts (see Section \ref{sec:prep_gal_cutouts} for details) from HST-ACS-WFC F814W unrotated tiles \citep{koekemoer2007} and generated a corresponding point-spread function (PSF) image for GALFIT (Section \ref{sec:galfit_psf}) using TinyTim \citep{krist2011}. To model the light profiles of galaxies with GALFIT as accurately as possible, for each galaxy, we adopted a case-specific model, taking into account the prior knowledge derived from the multiwavelength analysis presented in this study. We then estimated the initial parameters of that model through a close visual inspection of cutouts (see Sections \ref{sec:galfit_results_10913} to \ref{sec:galfit_results_4092}).

Our GALFIT analysis yields that the host of the WAT 10913 (the BGG of the group) is a giant elliptical galaxy. It is best fitted with a S\'ersic+PSF model (where PSF accounts for the AGN in the center), while simultaneously modeling the sky background, and another nearby source with another S\'ersic model (Section \ref{sec:galfit_results_10913}). The best-fit S\'ersic index for this galaxy is $n=3.825\pm0.006$ and the effective radius, $R_{e}\approx15\hspace{0.1cm}\mathrm{kpc}$, with the residual images revealing signs of an extended halo. Such a result is expected, as WATs typically correspond to BCGs or BGGs, which are often D or cD type galaxies \citep{burns1981}.

The host of the tailed radio galaxy 44, according to the best-fit model (S\'ersic+sky), is also a giant elliptical with an extended halo found in the residual, with S\'ersic parameters: $n=4.84\pm0.01$, $R_{e}\approx 8.8\hspace{0.1cm}\mathrm{kpc}$. This is consistent with its high stellar mass and brightness (see CMD and CSMD in Sections \ref{sec:cmd_analysis} and \ref{sec:csmd_analysis}).

The host of radio source 4092 has a close companion. We modeled both simultaneously (S\'ersic+S\'ersic+sky), and found the host of 4092 has a significant residual, notably in the form of spiral arms. We interpret this as evidence of a spiral structure induced in an originally elliptical galaxy via gravitational interaction with the companion (see Section \ref{sec:galfit_results_4092} for more details).

\subsection{Surface brightness profiles of host galaxies}
\label{sec:sbps_short}
\begin{figure}[!t]
\centering
\includegraphics[width=0.9\hsize]{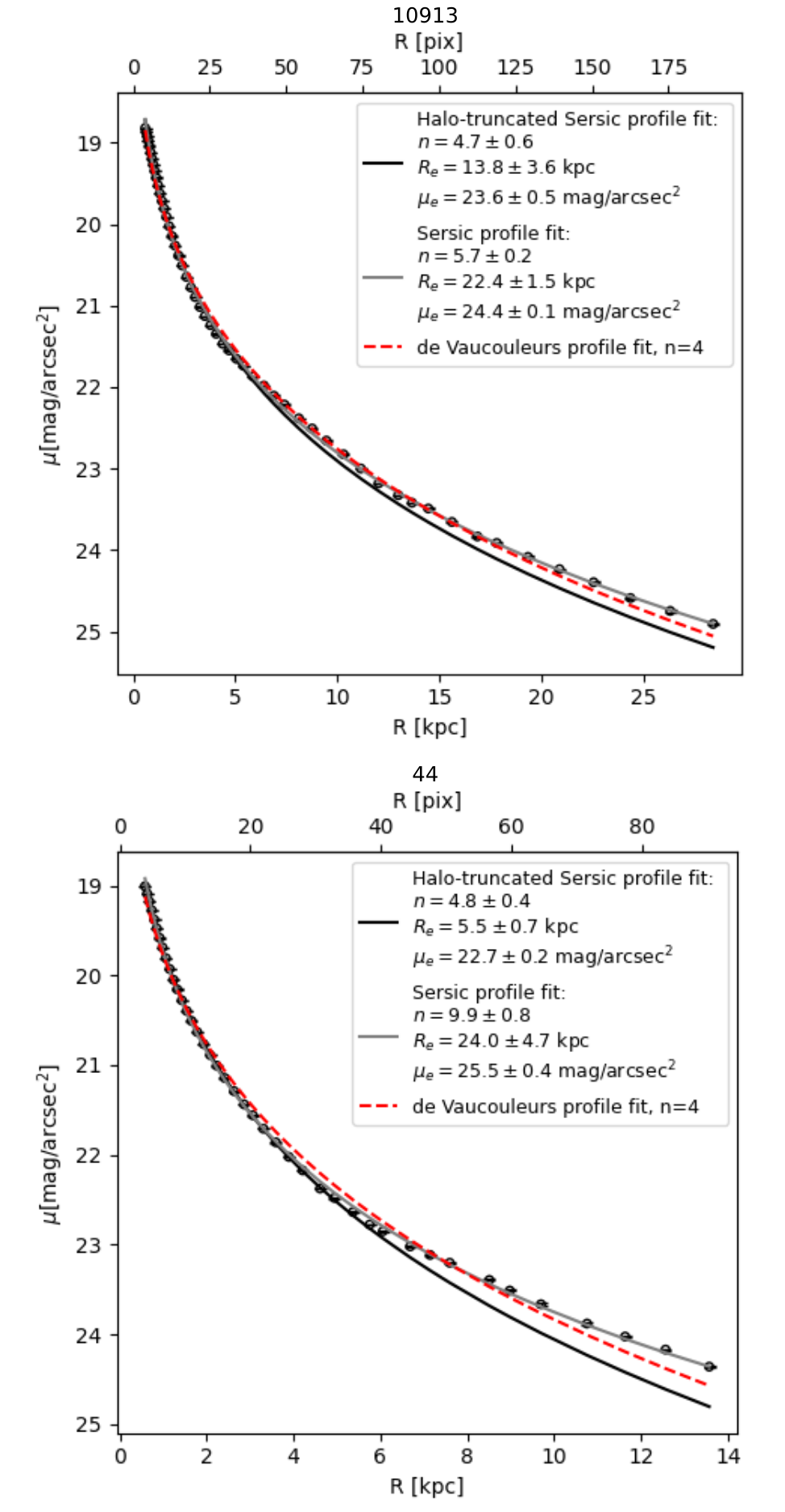}
\caption{
Surface brightness profiles from elliptical isophotal fits for hosts of radio galaxies 10913 (top panel) and 44 (bottom panel). In both panels, the best-fit S\'ersic model, fit to Halo-truncated data within $R\sim6\hspace{0.1cm}\mathrm{kpc}$ (for 10913) and $R\sim2.8\hspace{0.1cm}\mathrm{kpc}$ (for 44) is shown as a solid black line, with best-fit parameters reported in the legend. It closely follows the de Vaucouleurs profile (dashed red line). The S\'ersic fit on the full dataset is shown as a solid gray line. The upper x-axis represents the corresponding radius $R$ in pixels, and the y-axis is the ST magnitude per $\mathrm{arcsec^2}$.}
\label{Fig:10913_44_SBP}
\end{figure}

We used the Python \texttt{Photutils} package \citep{bradley25} to fit elliptical isophotes to HST-ACS F814W cutouts \citep{koekemoer2007} of host galaxies of radio galaxies 10913 and 44 to create the corresponding surface brightness profiles (SBPs). The fitting algorithm is the implementation of the iterative method introduced by \citet{jedrzejewski1987}, built to analyze elliptical systems. Therefore, we do not apply this method to the host galaxy of the point-like radio source 4092, given the GALFIT residual implies a non-elliptical and asymmetric geometry (see previous section). The process of isophote fitting is described in detail in appendix \ref{app:sbps}. The commonly used analytical form of S\'ersic profile is given in terms of equivalent radius, $R$. Therefore, from the results of the isophote fit, we created equivalent circular profiles $\mu(R)$. Here, $\mu$ is in units of $\mathrm{mag}/\mathrm{arcsec^2}$, magnitudes are in the Space Telescope (ST) magnitude system (defined to assign zero color to sources with constant flux per unit wavelength, and used in HST photometry packages), and $R$ is in kiloparsecs (kpc). For each galaxy, we plot its SBP $\mu(R)$, and fit S\'ersic model to the data points (see Fig.~\ref{Fig:10913_44_SBP}).
The S\'ersic profile is given as
\begin{equation}
\label{eq:sersic}
    I(R)=I_e\cdot e^{{b_n\bigg(\bigg(\frac{R}{R_e}\bigg)^{1/n}-1\bigg)}}
,\end{equation}
where $I$ is the intensity, $R_e$ the half-light radius, $I_e$  the intensity at $R_e$, $n$ is the S\'ersic index, and $b_n$ depends on $n$ according to
\begin{equation}
\label{eq:b_n}
b_n=2\hspace{0.05cm}n-\frac{1}{3}+\frac{4}{405\hspace{0.05cm}n}+\frac{46}{25515\hspace{0.05cm}n^2}+\frac{131}{1148175\hspace{0.05cm}n^3}-\frac{2194697}{30690717750\hspace{0.05cm}n^4}
.\end{equation}
The above expression (Equation \eqref{eq:sersic}) converted from an intensity to a surface brightness profile becomes
\begin{equation}
\label{eq:sersic_mu}
    \mu(R)=\mu_0+\frac{2.5}{\ln{10}}\cdot b_n\cdot \bigg(\bigg(\frac{R}{R_e}\bigg)^{1/n}-1\bigg)
,\end{equation}
where $\mu$ is surface brightness in units $\mathrm{mag}/\mathrm{arcsec^2}$, and $\mu_0$ is the surface brightness at $R_e$, which corresponds to
\begin{equation}
    \mu_0=-2.5\cdot \log_{10}{(I_e)}+\mathrm{ZP}+2.5\cdot \log_{10}{(\mathrm{PS^2})}
,\end{equation}
where $\mathrm{ZP}$ is the photometric zero point, and $\mathrm{PS}$ is the plate scale of the data.
We fit the model given by Equation \eqref{eq:sersic_mu} to the data and vary the three parameters $\mu_0$, $R_e$, and $n$. We also fit de Vaucouleurs model, a special case of S\'ersic model (Equation \eqref{eq:sersic_mu} for $n=4$), describing the radial distribution of surface brightness for elliptical galaxies. The results are presented in Fig.~\ref{Fig:10913_44_SBP}. For both galaxies, the best-fit S\'ersic model follows the de Vaucouleurs model very well except for larger radii, where we detect an excess of light. Based on this, and after a close inspection of the galaxy cutouts and GALFIT residual in DS9 \citep{DS9}, we propose that both galaxies have extended halos which start to dominate at $R\sim6\hspace{0.1cm}\mathrm{kpc}$ (10913), and $R\sim 2.8\hspace{0.1cm}\mathrm{kpc}$ (44). Therefore, we also perform a halo-truncated data fit, namely, a S\'ersic model fit to the data points up to these radii. This results in the following best-fit parameters: $n=4.7\pm 0.6$ and $R_e=13.8\pm3.6\hspace{0.1cm}\mathrm{kpc}$ for 10913, and $n=4.8\pm 0.4$, and $R_e=5.5\pm0.7\hspace{0.1cm}\mathrm{kpc}$ for 44. These values are in good agreement with the results of the GALFIT analysis, with the largest relative difference found to be $\sim38\%$ for $R_e$ for the host of radio galaxy 44. Any disagreement in results most probably originates from the differences in methods used. Unlike in an isophote fitting, in GALFIT analysis, the cutouts are convolved with a pregenerated PSF, additional nearby sources are simultaneously fitted away, instead of being masked and a minor difference may originate from the fact that any possible sky background is taken into account. In general, 2D GALFIT analysis is superior compared to 1D SBP fitting. However, disregarding the aforementioned differences, the results agree that both host galaxies have a steep inner region and a very extended outer wing with an excess of light compared to a standard elliptical galaxy's profile. The joint results of these two conducted analyses reveal that both host galaxies are giant ellipticals with extended halos.

\vspace{-0.3cm}
\section{Discussion}
\label{sec:discussion}
In Section \ref{sec:group_dyn_from_radiomorph}, we investigate the group's dynamical state using information on radio galaxies' morphology. In Section \ref{sec:unrelaxed_group}, we discuss the combined results of the multiwavelength analysis presented in this paper, placing them in the context of the group's dynamical state.
\vspace{-0.2cm}
\subsection{Group dynamics from radio morphology}
\label{sec:group_dyn_from_radiomorph}
\subsubsection{Jet bending:\ Interaction with the medium}
\label{sec:jet_bending_theory}
The jet bending observed in WATs near their core is believed to emerge due to the ram pressure exerted on the jets while the galaxy is moving at high velocity relative to the IGM (or ICM) of its host group \citep[or cluster,][]{begelman1979}. This velocity can be estimated from hydrodynamical models of jet bending. The jets can be described as a relativistic flow of plasma (see \citealt{odea2023} and references therein) obeying the relativistic and time-independent Euler's equation for an ideal fluid \citep{begelman1979, odea1985}, expressed as
\begin{equation}
\label{eq:begelman}
    \frac{\rho_j v_j^2 \gamma^2}{R}\approx \frac{\rho_{IGM} v_{gal/IGM}^2}{h}
,\end{equation}
where $\rho_j$ and $v_j$ are the jet density and bulk velocity, respectively, $R$ is the jet's radius of curvature, and $h$ the scale height, $\gamma=(1-\beta^2)^{-1/2}$ is the Lorentz factor, where $\beta=v_j/c$, $\rho_{IGM}$ is the IGM density, and $v_{gal/IGM}$ is the host galaxy's velocity relative to the IGM. The equation can be understood as describing the equilibrium between the ram pressure and the centrifugal force generated by the jet as it undergoes curvature.
In the next section, we estimate the velocity relative to the IGM of the BGG, which is the host of WAT galaxy 10913. Given the tailed radio morphology (i.e., the unresolved jets) of the other radio galaxy, 44, we are unable to estimate the true radius of curvature $R$ and to apply the above described approach in this case.
\vspace{-0.2cm}
\subsubsection{Group dynamics from the bending of WAT 10913}
\label{sec:bending_10913}
\begin{figure}[!t]
    \centering
    \includegraphics[width=0.48\textwidth]{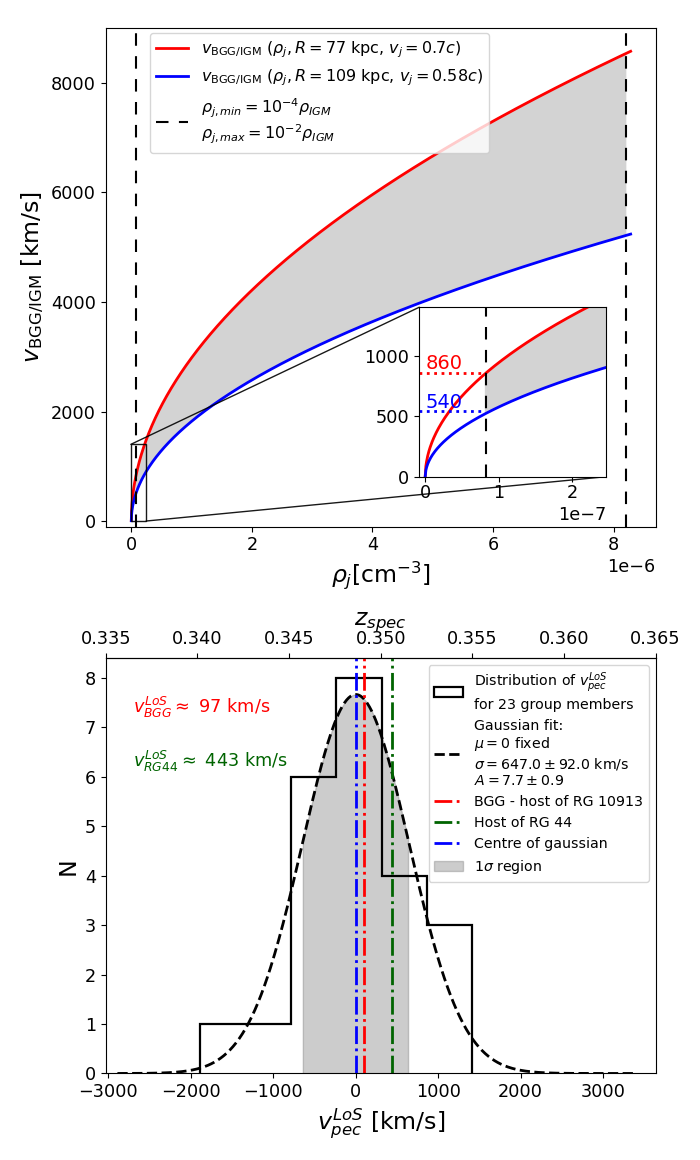}
    \caption{Top: Velocity of the BGG - host of WAT radio galaxy 10913, relative to the IGM as a function of the jet density in $\mathrm{cm}^{-3}$. The light-grey region denotes the range of possible values and solid curves represent limiting cases: $R=77\hspace{0.1cm}\mathrm{kpc}$, $v_j=0.7c$ (red), and $R=109\hspace{0.1cm}\mathrm{kpc}$, $v_j=0.58c$ (blue). Vertical dashed lines indicate jet density bounds: $\rho_{j,min}=10^{-4}\rho_{IGM}$ and $\rho_{j,max}=10^{-2}\rho_{IGM}$. In the lower right, the narrow range around $\rho_{j,min}$ is enlarged, revealing the minimum $v_{BGG/IGM}$ for the two limiting cases (red and blue color). Bottom: LoS peculiar velocities (redshifts) of the hosts of radio galaxies 10913 (BGG) and 44 compared to the (group) center of the Gaussian fit to 23 high confidence group members (see text for details).}
    \label{fig:v_gal_icm_rho_j_10913}%
\end{figure}

In case of WAT radio galaxy 10913, we observe the bending of jets close to the radio core.
We used Equation \eqref{eq:begelman} to place constraints on the velocity of its host galaxy (BGG) with respect to the IGM. 
To apply this model, which relies on non-projected quantities, we first analyzed the orientation of this radio galaxy. We place constraints on the two rotation angles ($\Theta\in[0^{\circ},45^{\circ}]$ and $\phi\in[32^{\circ},53^{\circ}]$, see appendix \ref{app:rotation} for details on angle definitions and visualization), which tilt the WAT radio structure out of the plane of the sky and introduce LoS component, thereby affecting how we observe it. This allows us to estimate the true (non-projected) radius of curvature, $R$. We did not correct the measured scale height parameter, $h$, for projection effects as its impact on the final result is minor relative to the uncertainties associated with the other parameters. Both parameters $R$ and $h$ were estimated at the bending point, which we assumed to correspond to the location of a hotspot found along the northern jet slightly away from the core (see panel h in Fig.~\ref{fig:rotation_problem}).
We estimated a scale height, $h\sim 6.5\hspace{0.1cm}\mathrm{kpc}$, and a range for projection-corrected radius of curvature from $R\sim 77\hspace{0.1cm}\mathrm{kpc}$ (for $\Theta=0$, $\phi=32^{\circ}$) to $R\sim109\hspace{0.1cm}\mathrm{kpc}$ (for $\Theta=45^{\circ}$, $\phi=53^{\circ}$). \citet{jetha2006} reported bulk jet velocities in WATs between $0.3\hspace{0.1cm}\mathrm{c}$ and $0.7\hspace{0.1cm}\mathrm{c}$. Given that we observe hotspots in the lobes of 10913 and assuming these are jet-termination shocks, the velocity should be larger or equal to the internal sound speed for a relativistic plasma $\approx 0.58\hspace{0.1cm}\mathrm{c}$ \citep{jetha2006}. Therefore, we assumed a possible range of bulk jet velocities from $0.58\hspace{0.1cm}\mathrm{c}$ to $0.7\hspace{0.1cm}\mathrm{c}$. We used the density of the IGM based on X-ray observations (Section \ref{sec:xray_analysis}). As the jet density, $\rho_j$, is not well constrained, we expressed $v_{BGG/IGM}$ as a function of it, assuming a range of jet densities from $10^{-4}\rho_{IGM}$ to $10^{-2}\rho_{IGM}$, which is widely used in hydrodynamical simulations \citep{rossi2004}. The dependence is shown in the top panel of Fig.~\ref{fig:v_gal_icm_rho_j_10913}. The gray shaded region represents the range of possible solutions, the width of which is affected by uncertainties in radio morphology parameters (projection effects), $v_j$ and $\rho_j$. For the lowest possible jet density and bulk velocity ($\rho_j=10^{-4}\rho_{IGM}$, $v_j=0.58\hspace{0.1cm}\mathrm{c}$), and the largest possible curvature radius ($R\sim109\hspace{0.1cm}\mathrm{kpc}$), we find $v_{BGG/IGM}$ to be $\sim 540\hspace{0.1cm}\mathrm{km/s}$. We note that for any change in the above parameters and within the possible ranges, this value would further increase, namely $v_{BGG/IGM}\gtrsim540\hspace{0.1cm}\mathrm{km/s}$.\par
The high relative velocity $v_{BGG/IGM}$ found is in agreement with the requirements of the jet bending theory. We compare this velocity with the BGG's peculiar velocity. We find the LoS component of the latter $v_{BGG}^{LoS}\sim 97\hspace{0.1cm}\mathrm{km/s}$ (in the direction away from the group's center) from the difference between the spectroscopic redshift of the BGG $z_{BGG}^{spec}$, and the center of the Gaussian distribution of high-confidence spectroscopic redshifts for 23 most probable group members $\mu=0.3486\pm0.0005$ (see Section \ref{sec:cmd_galaxysampleandphotometry} for details about this sample), using ~$v_{BGG}^{LoS}=c(z_{BGG}^{spec}-\mu)/(1+\mu)$ \citep{Davis2014}. The distribution of members' LoS peculiar velocities, calculated using the above formula, and the corresponding Gaussian fit is shown in the bottom panel of Fig.~\ref{fig:v_gal_icm_rho_j_10913}, where the blue and red dot-dashed lines represent the Gaussian center and $v_{BGG}^{LoS}$, respectively. The latter fit was performed with the Gaussian center fixed at 0, corresponding to the expected value in the group’s rest frame. The number of histogram bins used for the Gaussian fit was determined with the \texttt{auto} option in Python, which automatically finds the optimal bin number for a given dataset as the maximum of the Sturges and Freedman-Diaconis bin choice. We ruled out selection bias by comparing the normalized distribution of spectroscopic redshifts with the one of photometric redshifts, and with the complete sample which includes both types. All distributions match, implying we can safely assume the spectroscopic sample accurately represents the LoS motions of the group galaxies. Using constraints we placed on the rotation angle $\phi$, we estimate the full length and the direction of the BGG peculiar velocity vector, namely, $v_{BGG}$ is between $\sim114\hspace{0.1cm}\mathrm{km/s}$ (for $\phi=32^{\circ}$) and $\sim 160\hspace{0.1cm}\mathrm{km/s}$ (for $\phi=53^{\circ}$), in the direction towards the west and away from the observer tilted at the corresponding angle $\phi$ from the plane of the sky. In determining the full direction of $v_{BGG}$, the other rotation angle $\Theta$ is of no importance (see the angle definition in appendix \ref{app:rotation}), and the third angle of rotation of the WAT structure, namely, the one in the plane of sky was set to 0 (i.e., the peculiar velocity component in the plane of the sky is purely towards the west). The latter is a reasonable value based on the observations of the radio structure (see Fig.~\ref{fig:10913_3ghz_HST} in Section \ref{sec:radio_src_and_hosts} and panel g of Fig.~\ref{fig:rotation_problem}), and any potential small deviations from it do not affect the above derived velocities $v_{BGG}^{LoS}$ and $v_{BGG}$.\par
The low peculiar velocity is consistent with what is commonly observed in relaxed systems \citep{beers1995,Ow2001,lauer2014}. However, such a result yields a large difference between the BGG peculiar velocity and its velocity with respect to the IGM derived above. This implies the large relative velocity we find is mostly due the rapid bulk motion of the IGM itself (i.e., $v_{IGM}\gtrsim380\hspace{0.1cm}\mathrm{km/s}$) in the direction opposite to the peculiar movement of the BGG, suggesting an early-stage group merger scenario.

\vspace{-0.3cm}
\subsubsection{Radio Galaxy 44: Origin of the tailed morphology}
\label{sec:headtail_or_wat_44}
The radio morphology of galaxy 44 implies two possible scenarios. It could be this is a radio galaxy inclined at a very large angle $\Theta$ with respect to the plane of the sky. Then, the galaxy would experience a strong relativistic beaming effect, diminishing the flux of the receding jet, keeping it under the detection limit. However, this scenario is less plausible, since relativistic beaming is generally expected to be weak in FRI sources, where jets are launched at relativistic velocities, but become subrelativistic a few kpc away from the core \citep{Laing2002a,Laing2002b}.\par
Alternatively, this might be a HT galaxy (extreme and unresolved case of a NAT), requiring very large velocities with respect to the IGM ($\gtrsim10^3\hspace{0.1cm}\mathrm{km/s}$, \citealt{miley1972}), and/or a very dense IGM at the galaxy's location. We find the latter, namely, the electron density for the group's core (encompassing the galaxy's location) in Section \ref{sec:xray_analysis}. The estimated value, $n_e = (8.2 \pm 0.3) \times 10^{-4}\hspace{0.1cm}\mathrm{cm}^{-3}$, lies at the lower end of the range typically reported in the literature. The characteristic values for cluster cores are generally around $\sim10^{-3}\hspace{0.1cm}\mathrm{cm}^{-3}$ \citep{Haarsma2010,Bohringer2016}, with cool cores in massive clusters reaching densities of $\sim10^{-2}$ to $10^{-1}\ \mathrm{cm}^{-3}$ \citep{Hudson2010}, and poor clusters, galaxy groups, and dynamically young systems often exhibiting lower densities, on the order of $\sim10^{-4}\ \mathrm{cm}^{-3}$ \citep{Bubul2024}. Therefore, it is not likely a high IGM density would be responsible for extreme jet bending in our galaxy.\par
Multiple studies of NATs (and HTs) in groups and clusters have uncovered high LoS peculiar (and full) velocities for their hosts; namely, $v_{LoS}\sim500-3700\hspace{0.1cm}\mathrm{km/s}$ (\citealt{sebastian2017} in a study of seven NATs; \citealt{bruno2024}, \citealt{bushi2025}). They generally find velocities to be on the same  order (or a few times larger) as the corresponding LoS cluster dispersions. Here, we find $v_{RG44}^{LoS}\sim443\hspace{0.1cm}\mathrm{km/s}$ (Fig.~\ref{fig:v_gal_icm_rho_j_10913}). This is about two-thirds of the LoS dispersion found from the Gaussian fit (Fig.~\ref{fig:v_gal_icm_rho_j_10913}) for our massive group, $\sigma=647\pm92\hspace{0.1cm}\mathrm{km/s}$, and it differs somewhat from the above expectation for NATs. A very large plane of the sky peculiar velocity or a fast bulk IGM motion are still plausible explanations; however, given the lack of information on the true 3D peculiar velocity, we have not been able to investigate this further.
Based on the above, it seems to be more likely that radio galaxy 44 is an unresolved NAT, that is, a HT galaxy. Higher resolution radio imaging would be required to confirm this interpretation and resolve its detailed morphology. To avoid any overinterpretation, we have referred to radio galaxy 44 throughout this paper simply as a tailed radio galaxy.
\vspace{-0.2cm}
\subsection{Evidence for an unrelaxed galaxy group}
\label{sec:unrelaxed_group}
The combined results of the above presented multiwavelength study imply this group is an unrelaxed system - possibly a merger of smaller structures (subgroups) of galaxies.
We find an asymmetric, irregular distribution of the IGM, typical of unrelaxed systems \citep{bird94}, and our analysis of the X-ray spectra reveals a high temperature of the medium of $2.4\pm0.6$~keV and an electron density of $(8.2\pm0.3) \times 10^{-4}\hspace{0.1cm}\mathrm{cm^{-3}}$. This temperature is consistent with the values of \citet{Kettula}, who also model the X-ray spectra, and is higher (albeit only marginally) than the value found from $L_X-T_X$ scaling by \citet{Gozaliasl}.
The group is very massive, with $M_{200}$ from $L_X-M_{200}$ scaling \citep{Gozaliasl} $\approx 320\%$ above the median for X-ray detected galaxy groups in the redshift bin [0.1,0.5] in the COSMOS field, and weak lensing $M_{200}=0.89\substack{+0.50 \\ -0.41}\times 10^{14}M_{\odot}$ \citep{Kettula}.\par
We found seven radio sources within the group. Two of them are tailed radio galaxies, with one of them WAT radio galaxy (10913) and the other one tailed radio galaxy (44). WAT (and bent) radio morphology itself is indicative of unrelaxed systems \citep{smolcic2007, oklopcic2010}. We find both radio galaxies are hosted by giant, bright, massive elliptical galaxies with extended halos, with the host of 10913 corresponding to the BGG, and most massive galaxy, and the host of 44 the second brightest, and second most massive.\par
Our VT analysis detected a group-related overdensity of galaxies in this area, and the follow-up color-magnitude analysis revealed the domination of red galaxies among them, typical for galaxy groups and clusters \citep{Dressler}. For a virialized group, a regularly shaped overdensity would be expected with the BGG in the centre, dominating the dynamics, however, the overdensity we detect is irregular, and both the BGG and the second brightest galaxy seem to be misplaced from its center suggesting a non-virialized system. \citet{bird94} showed that large peculiar velocity is strongly correlated with the existence of substructure in galaxy clusters. We find weak signs of galaxy substructuring (subgrouping) in our system, where the detected subgroups are found around each of the point radio sources: 4092, 10366, and 1549. An exception to the weak substructuring is a very dense, elongated accumulation of galaxies found in the far east of the group-related overdensity. The relatively weak presence of substructure agrees with the low value we find for peculiar velocity of the BGG $\sim 114-160\hspace{0.1cm}\mathrm{km/s}$. The conclusions about the shape of the detected overdensity and the absence of strong substructuring are possibly affected by the lack of data, namely, galaxies with valid photometric redshifts in masked regions and generally a lack of reliable spectroscopic redshifts. The latter could have also affected the calculated peculiar velocity. Nevertheless, based on all of the above arguments, we suggest this system could be a merger of two (or more) smaller (less massive) galaxy structures, possibly individual groups. This scenario is further supported by our finding of a large velocity of the BGG relative to the IGM: $v_{BGG/IGM}\gtrsim 540\hspace{0.1cm}\mathrm{km/s}$. This, combined with the low peculiar velocity we find, implies a large velocity of the IGM: $v_{IGM}\gtrsim380\hspace{0.1cm}\mathrm{km/s}$, namely, the WAT bending in 10913 mostly originates from the movement of the IGM itself. Such a result suggests the potential group merger is in an early state.\par
The aforementioned accumulations of galaxies around radio sources 4092, 10366, and 1549, could possibly correspond to small sub-groups. If this is the case, it is possible these may be infalling into the gravitational potential of the larger structure dominated by the host of WAT galaxy 10913. The host of radio source 633 lies outside the VT detected group-related overdensity. Combined with its spectroscopic redshift, this suggests association with a separate galaxy structure at lower redshift ($z \approx 0.3$). Given the increased galaxy density around radio source 4092, it is not surprising that we have found its host galaxy in a close gravitational interaction with a companion, thereby serving as an example of a galaxy merger within a galaxy group merger.

\section{Summary}
\label{sec:summary}
In this work, we performed a multiwavelength study of a massive galaxy group in the COSMOS field at redshift $z=0.349$. The key findings are summarized below.\begin{enumerate}
    \item Irregularly shaped diffuse X-ray emission from the IGM was detected with a high temperature $T_{X}=2.4\pm0.6$~keV and density $n_e=(8.2\pm0.3) \times 10^{-4}\hspace{0.1cm}\mathrm{cm^{-3}}$. The group is very massive, luminous, and hot compared to other groups in the COSMOS at these redshifts.
    \item A total of seven radio sources were found within the galaxy group. Two of them are tailed radio galaxies, both hosted by giant ellipticals with extended halos. One is a WAT, corresponding to the BGG, which is the most massive, and the other one is tailed radio galaxy that is the second-brightest and the second most massive group galaxy.
    \item The VT analysis revealed an irregularly shaped group-related overdensity of optical galaxies. Radio galaxies' hosts are found on the outskirts of this overdensity. The low LoS peculiar velocity found for the BGG is consistent with the absence of strong substructuring detected in the galaxy distribution, except for a few minor accumulations, possibly small individual groups merging with the one around the main WAT and an extremely dense, elongated substructure in the far east of the group.
    \item We found the high velocity of the BGG relative to the IGM, $\gtrsim540\hspace{0.1cm}\mathrm{km/s}$, mainly originating from the bulk IGM movement. The combined results suggest the system is unrelaxed, that is, a possible early-stage galaxy group merger. The results are consistent with WATs serving as an indication of galaxy clustering and suggesting dynamically young systems.
    \end{enumerate}
In conclusion, our analysis reveals a massive, dynamically young galaxy group in the early stages of assembly, where the interaction between the galaxies and the IGM gives rise to tailed radio morphologies. The study confirms that WATs can serve as effective tracers of the dynamical processes shaping galaxy groups and clusters. Future deep, high-resolution multiwavelength observations, combined with detailed hydrodynamical modeling, would be crucial to constrain the underlying physical mechanisms and to further investigate this system, as well as the general link between tailed morphology and the dynamical properties (and growth) of galaxy assemblies.
\begin{acknowledgements}
The authors thank K. Virolainen for her valuable assistance in rendering the RGB composite image. 
ID acknowledges funding by the European Union – NextGenerationEU, RRF M4C2 1.1, Project 2022JZJBHM: "AGN-sCAN: zooming-in on the AGN-galaxy connection since the cosmic noon" - CUP C53D23001120006.\\
This research made use of \texttt{Photutils}, an \texttt{Astropy} package for
detection and photometry of astronomical sources \citep{bradley25}.\\
This research has been supported by the European Regional Development Fund under grant agreement PK.1.1.10.0007 (DATACROSS)\\
This work was supported by the project “Implementation of cutting-edge research and its application as part of the Scientific Center of Excellence for Quantum and Complex Systems, and Representations of Lie Algebras“, Grant No. PK.1.1.10.0004, co-financed by the European Union through the European Regional Development Fund - Competitiveness and Cohesion Programme 2021-2027.
\end{acknowledgements}
\vspace{-0.5cm}
\bibliographystyle{aa}
\bibliography{bibliography}

\begin{appendix}
\onecolumn

\section{Radio and X-ray sources within the group and its properties}
\label{app:tables}
\vspace{-0.3cm}
\begin{table*}[!h]
\caption{Radio sources found within the group, presented together with their multiwavelength properties (counterparts within $3^{\prime\prime}$ radius) available in the literature (see notes below).}
\label{table:srces}      
\centering
\scalebox{0.8}{
\begin{tabular}{l|c|c|c|c|c|c|c}
\hline\hline
     & 10913 & 44 & 4092 & 1549 & 10366 & 633 & 528 \\ \hline
    Ra & 150.1179 & 150.1104 & 150.1461 & 150.1080 & 150.1386 & 150.1602 & 150.1235\\
    Dec & 2.6843 & 2.7083 & 2.6985 & 2.6955 & 2.7030 & 2.6993 & 2.6717\\
    \begin{tabular}{@{}l@{}}Spectroscopic\\R edshift\tablefootmark{$\dagger$}\end{tabular} & 0.3491 & 0.3506 & 0.3505 & 0.3492 & 0.354 & 0.3044 & 0.345 \\ \hline
    \multicolumn{8}{c}{\centering Radio properties \hspace{0.5cm}(Detected\hspace{0.1cm}/\hspace{0.1cm}Flux density [mJy]\hspace{0.1cm}/\hspace{0.1cm}Multi)}\\ \hline
   3 GHz (VLA) & \begin{tabular}{@{}c@{}}YES\hspace{0.1cm}/\hspace{0.1cm}\\$32.09$\hspace{0.1cm}/\\ \hspace{0.1cm}1\end{tabular} & \begin{tabular}{@{}c@{}}YES\hspace{0.1cm}/\hspace{0.1cm}\\$2.3\pm0.1$\hspace{0.1cm}/\\ \hspace{0.1cm}0\end{tabular} & \begin{tabular}{@{}c@{}}YES\hspace{0.1cm}/\hspace{0.1cm}\\$0.028\pm0.003$\hspace{0.1cm}/\\ \hspace{0.1cm}0\end{tabular} & \begin{tabular}{@{}c@{}}YES\hspace{0.1cm}/\hspace{0.1cm}\\$0.071\pm0.004$\hspace{0.1cm}/\\ \hspace{0.1cm}0\end{tabular} &
   \begin{tabular}{@{}c@{}}YES\hspace{0.1cm}/\hspace{0.1cm}\\$0.011\pm0.002$\hspace{0.1cm}/\\ \hspace{0.1cm}0\end{tabular} & 
   \begin{tabular}{@{}c@{}}YES\hspace{0.1cm}/\hspace{0.1cm}\\$0.121\pm0.007$\hspace{0.1cm}/\\ \hspace{0.1cm}0\end{tabular} &
   \begin{tabular}{@{}c@{}}YES\hspace{0.1cm}/\hspace{0.1cm}\\$0.152\pm0.008$\hspace{0.1cm}/\\ \hspace{0.1cm}0\end{tabular} \\ \hline
   \multicolumn{8}{c}{\centering Xray properties \hspace{0.5cm} (Detected\hspace{0.2cm}/\hspace{0.2cm}Full X-ray (2-10 keV) Flux\tablefootmark{$\star$} [erg/s/$\mathrm{cm^{2}}$])}\\ \hline
   Chandra (2016)  & YES\hspace{0.1cm}/\hspace{0.1cm}$3.15\times 10^{-15}$ & YES\hspace{0.1cm}/\hspace{0.1cm}$4.6\times 10^{-15}$ & NO\hspace{0.1cm}/\hspace{0.1cm}- & NO\hspace{0.1cm}/\hspace{0.1cm}- & NO\hspace{0.1cm}/\hspace{0.1cm}- & NO\hspace{0.1cm}/\hspace{0.1cm}- & YES\hspace{0.1cm}/\hspace{0.1cm}$3.93\times 10^{-15}$ \\ \hline
   \multicolumn{8}{c}{\centering Photometric properties of optical counterparts}\\ \hline
   CLASSIC ID & - & 1358068 & 1352485 & 1346601 & 1351402 & 1353634 & - \\
   HSC $i$ mag Aper2 & - & $19.322\pm0.001$ & $19.690\pm0.002$ & $19.944\pm0.002$ & $21.764\pm0.004$ & $19.815\pm0.002$ & - \\
   COSMOS2015 ID & 901584 & 923481 &  921175 & 920711 & 925241 & 923235 & 902831 \\
   SC $i^+$ mag Aper2 & $18.826\pm0.002$ & $19.391\pm0.003$ & $19.742\pm0.004$ & $19.991\pm0.003$ & $21.83\pm0.01$& $19.883\pm0.004$ & $19.922\pm0.004$\\ 
   ACS F814W mag & $17.310\pm0.001$ & $18.350 \pm0.001$ & $19.330\pm0.001$ & $19.360\pm0.001$ & $21.310\pm0.007$& $18.990\pm0.002$ & $19.050\pm0.002$\\
 \hline
\end{tabular}
}
\tablefoot{First two, 10913 and 44, are tailed radio galaxies. The radio data were adopted from the source catalog at 3 GHz \citep{3ghz_smol_b}, X-ray properties are from Chandra catalog of point sources \citep{civano2016}, and optical/NIR photometry data from CLASSIC \citep{COSMOS2020} and COSMOS2015 \citep{Laigle} multiwavelength photometry catalogs. HSC $i$ band magnitudes were adopted from CLASSIC, and SC $i^{+}$ and HST-ACS-F814W magnitudes from COSMOS2015 catalog. Only selected (relevant) properties are presented here, and for the others one should refer to the corresponding catalogs. Errors are provided when available.\\
\tablefoottext{$\dagger$}{From \citet{Khostovan2025}.}\\
\tablefoottext{$\star$}{For details on X-ray detections (full/soft/hard) see the analysis of X-ray point sources within the group, Table~\ref{table:xray_sources}.}
}
\end{table*}

\begin{table*}[!h]
    \caption{Properties of the galaxy group adopted from the literature (see notes below).}
    \centering
    \begin{tabular}{c|c|c|c|c|c|c|c}
    \hline\hline
    \multicolumn{8}{c}{COSMOS X-Ray group catalog \citep{Gozaliasl}} \\ \hline
       ID  & Ra & Dec & Redshift & $R_{200}$\tablefootmark{a} [$^{\prime\prime}$] & $M_{200}$\tablefootmark{b} [$\mathrm{M_{\odot}}$] & $L_X$\tablefootmark{c} [erg/s] & $T_X$\tablefootmark{d} [keV] \\ \hline
       10237  & 150.11756
 & 2.69252 & 0.349 & 168 & $(9.6\pm0.3)\times 10^{13}$ & $(1.63\pm0.08)\times10^{43}$ & $1.68\pm0.04$ \\ \hline
 \multicolumn{8}{c}{Study of \citet{Kettula}}\\ \hline
 \multicolumn{2}{c|}{X-ray analysis} & \multicolumn{6}{c}{Weak lensing analysis} \\ \hline
 \multicolumn{2}{c|}{$T_X$\tablefootmark{e} [keV]} & \multicolumn{3}{c|}{$M_{500}$\tablefootmark{f} 
 $[10^{14} \mathrm{M_{\odot}}]$} & \multicolumn{3}{c}{$M_{200}$\tablefootmark{g} $[10^{14}\mathrm{M_{\odot}}]$} \\ \hline
 \multicolumn{2}{c|}{$2.2\substack{+2.1 \\ -0.5}$} & \multicolumn{3}{c|}{$0.64\substack{+0.36 \\ -0.29}$} & \multicolumn{3}{c}{$0.89\substack{+0.50 \\ -0.41}$} \\
 \hline
\end{tabular}
\tablefoot{
Properties are taken from the COSMOS X-ray group catalog \citep{Gozaliasl} and the results of \citet{Kettula}. For details on the derivation of the presented parameters, see the corresponding papers. Presented values are rescaled to the cosmology we use in this paper (if
needed).\\
\tablefoottext{a}{Group's radius where the internal density of halo is 200 times the critical density of the universe}\\
\tablefoottext{b}{Group's total mass within $R_{200}$ from $L_X-M_{200}$ scaling}\\
\tablefoottext{c}{Total rest-frame X-ray luminosity ($0.1-2.5\hspace{0.1cm}\mathrm{keV}$) within $R_{500}$}\\
\tablefoottext{d}{IGM temperature from $L_X-T_X$ scaling}\\
\tablefoottext{e}{IGM temperature from modeling X-ray spectra}\\
\tablefoottext{f}{Weak lensing group's mass within $R_{500}$ centered on X-ray peak}\\
\tablefoottext{g}{Weak lensing group's mass within $R_{200}$ centered on X-ray peak}
}
\label{table:xray_group_properties}
\end{table*}

\begin{table*}[!h]
\caption{Chandra-detected X-ray point sources within the galaxy group (Section \ref{sec:xray_ps_within_the_group}) sorted by full 0.5-10 keV X-ray flux in descending order.}         
\label{table:xray_sources}      
\centering
\scalebox{1}{
\begin{tabular}{c c c c c c }
\hline\hline                      
ID & Ra & Dec & \begin{tabular}{@{}c@{}} flux\_f (0.5-10 keV) \\ $10^{-15}[\mathrm{erg/s/cm^2}]$ \end{tabular}  & \begin{tabular}{@{}c@{}}flux\_s (0.5-2 keV)\\ $10^{-16}[\mathrm{erg/s/cm^2}]$\end{tabular} & \begin{tabular}{@{}c@{}}flux\_h (2-10 keV) \\ $10^{-15}[\mathrm{erg/s/cm^2}]$\end{tabular} \\
\hline                    
   lid\_2045 & 150.110667 & 2.708574 & 4.6 & 14.4 & \textbackslash \\  
   lid\_183 & 150.123608 & 2.671742  & 3.93 & 6.67 & 3.73 \\
   lid\_182 & 150.117955 & 2.684619  & 3.15 & 9.47 & 2.11 \\
   lid\_3144 & 150.15028 &  2.691248 & 2.26 & \textbackslash & 2.32 \\
   lid\_3137 & 150.11982 & 2.709775  & 2.15 & \textbackslash & 3.37 \\
\hline                  
\end{tabular}
}
\tablefoot{
The first and the third X-ray sources correspond to the tailed radio galaxies 44 and 10913 (WAT), respectively. The second one corresponds to point radio source 528, also found within the group.
}
\end{table*}
\clearpage

\twocolumn
\section{Flux density comparison}
\label{app:compare_flux}
We compared the 3 GHz flux densities calculated for the two radio galaxies 10913 and 44 in Section \ref{sec:radiomorph_and_lum} with the catalog values \citep{3ghz_smol_a}. The flux densities are consistent for 10913 and marginally so for 44, as shown in Fig.~\ref{Fig:compare_flux}.
\begin{figure}[!h]
\centering
\includegraphics[width=0.4\textwidth]{comparison_with_catalog_fluxes.png}
\caption{Comparison of 3 GHz radio flux densities calculated in this paper for the two tailed radio galaxies 10913 and 44 (Section \ref{sec:radiomorph_and_lum}) and flux densities from VLA 3 GHz catalog \citep{3ghz_smol_a}.}
\label{Fig:compare_flux}
\end{figure}

\vspace{-0.5cm}
\section{GALFIT analysis of host galaxies of radio sources}
\label{app:Galfit}
\subsection{Preparing galaxy cutouts}
\label{sec:prep_gal_cutouts}
We extracted the galaxy images from the HST-ACS-WFC unrotated tiles observed in the F814W band (\citealt{koekemoer2007}, also see Section \ref{sec:opticalnirdata} of this paper). We choose the unrotated tiles over the mosaic to simplify the PSF matching procedure given the tiles are in the default unrotated frame of the ACS-WFC CCDs.
For each galaxy, the corresponding cutout is centered at the galaxy’s Ra and Dec coordinates (wrt. Table~\ref{table:srces}). The cutout dimensions were chosen to cover all the galaxy's light, as well as a portion of the sky, while trying to exclude additional sources. However, neighboring objects remain in certain cases, as further truncation would have removed significant portions of the galaxy's light. Dealing with the remaining objects when performing GALFIT analysis is described in more detail for individual cases (if applicable) in Sections \ref{sec:galfit_results_10913} to \ref{sec:galfit_results_4092}.
In each cutout, we multiply the pixel values by the EXPTIME adopted from the corresponding header (EXPTIME=4056 s for all cutouts) to convert to analog-to-digital units (ADUs, electrons in this case), as required by GALFIT \citep{peng2010} to generate the sigma image internally.

\vspace{-0.3cm}
\subsection{Preparing a PSF image}
\label{sec:galfit_psf}
For an accurate modeling of a light distribution of a galaxy, GALFIT needs a PSF image describing the two-dimensional distribution of
light in the telescope focal plane for an astronomical point source. GALFIT convolves the cutout (or each of its parts if the convolution
kernel size is smaller than the image itself) with the PSF image before fitting. Here we set the kernel size to (125 pix, 125 pix) which is smaller than the cutout size for all three galaxies. The TinyTim tool \citep{krist2011} is used to generate the PSF image in three steps: tiny1, tiny2, and tiny3. In tiny1, we input details about the camera, detector, filter, and PSF expectations. These parameters remain the same for all galaxies (see Table~\ref{table:tinytim}). Tiny1 produces a parameter file, which tiny2 uses to create the PSF function. Since ACS data are significantly distorted, tiny3 integrates the PSF onto a distorted pixel grid. Without additional sampling, the PSF would have the ACS-WFC default plate scale of $(\mathrm{dx, dy})=(0.050, 0.050)\hspace{0.1cm}^{\prime\prime}\mathrm{/pixel}$, while the processed data \citep{koekemoer2007} have a $(\mathrm{dx, dy})=(0.030, 0.030)\hspace{0.1cm}^{\prime\prime}\mathrm{/pixel}$ scale. To properly match GALFIT’s requirements, we generate a 5× oversampled PSF (plate scale $(\mathrm{dx, dy})=(0.010, 0.010)\hspace{0.1cm}^{\prime\prime}\mathrm{/pixel}$) using SUB = 5 in tiny3. The PSF Fine-Sampling Factor in GALFIT is then set to 3.
\begin{table}[]
\caption{Input parameters for tiny1 (TinyTim) to create the PSF image for the GALFIT analysis.}
\label{table:tinytim}
\vspace{-0.7cm}
\begin{center}
\resizebox{\linewidth}{!}{
\begin{tabular}{c|c}
\hline\hline
Parameter & Value    \cr
\hline  
Camera    & ACS - Wide Field Channel \\ \hline
Detector chip (1 or 2)& 2  \\ \hline 
Position on the detector [pixels] & 2073, 1035  \\ \hline
Filter & f814 \\ \hline
Form of an object spectrum & M3V \\ \hline
PSF diameter [$^{\prime\prime}$] & $0.242$ ($0.095^{\prime\prime}$ FWHM PSF cut at $3\sigma$) \\ \hline
Secondary mirror despace [microns] & -3 (as in \citealt{HA2017})\\
\hline  
\end{tabular}
}
\tablefoot{The value of secondary mirror despace was set to the typical value according to \citet{HA2017}.}
\end{center}
\vspace{-0.4cm}
\end{table}

\vspace{-0.3cm}
\subsection{10913: Model, initial parameters, and results}
\label{sec:galfit_results_10913}
\begin{figure*}[!t]
\centering
\includegraphics[width=0.79\textwidth]{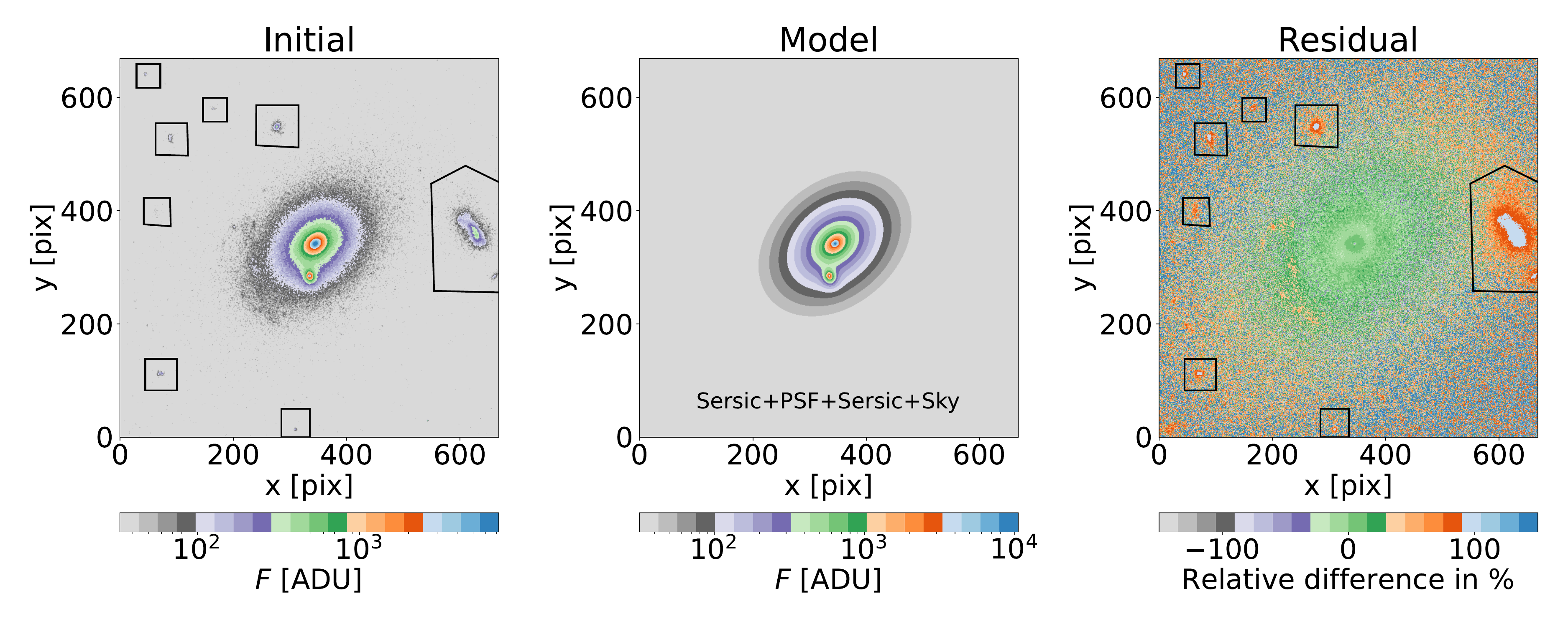}
\caption{GALFIT analysis results for the host galaxy of radio galaxy 10913 (initial, model, and residual images). The initial and model images are shown in $\log_{10}$ scale, while the residual image is in linear scale. The model is S\'ersic+PSF+S\'ersic (for another source)+Sky. GALFIT finds S\'ersic index and the effective radius of the host galaxy to be $n=3.825\pm0.006$ and $R_{e}=102.8\pm0.3\hspace{0.1cm}\mathrm{pix}$ ($\approx15\hspace{0.1cm}\mathrm{kpc}$). Black polygons represent the masked areas.}
\label{Fig:10913_galfit}
\end{figure*}

We model the host galaxy of radio galaxy 10913 with S\'ersic profile plus a PSF in the center for the corresponding AGN. This combined model was shown to be the best fit for the most of X-ray detected AGN host galaxies in the COSMOS field \citep{dewsnap2023}. There is an additional nearby optical source in the cutout which is not detected in the \textit{ACS I-band Photometry Catalog} of optical sources \citep{leauthaud2007}, most probably due to the combination of its low brightness and proximity to the optical counterpart of radio galaxy 10913. Masking this source would result in removing a large portion of our galaxy's light as well. Therefore, we simultaneously fit this source with another S\'ersic profile. We mask all other additional sources. We also let GALFIT to simultaneously fit the sky background. This combined S\'ersic+PSF+S\'ersic+sky model has proved to be very successful in fitting away the light in the cutout with the reduced $\chi^{2}$ value of $\sim0.7$. The initial parameters of the model were estimated through a close inspection of the sources in DS9 \citep{DS9}. GALFIT finds the S\'ersic index value of the host galaxy $n=3.825\pm0.006$, and the effective radius $R_{e}=102.8\pm0.3\hspace{0.1cm}\mathrm{pix}$, corresponding to $R_{e}\approx15\hspace{0.1cm}\mathrm{kpc}$ (using spectroscopic redshift). The full list of the GALFIT output model parameters can be found in Table~\ref{table:galfit_output_10913}. The resulting GALFIT cube (initial, model, and residual image) is shown in Fig.~\ref{Fig:10913_galfit}.

\vspace{-0.3cm}
\subsection{44: Model, initial parameters and results}
\label{sec:galfit_results_44}
\begin{figure*}
\centering
\includegraphics[width=0.79\textwidth]{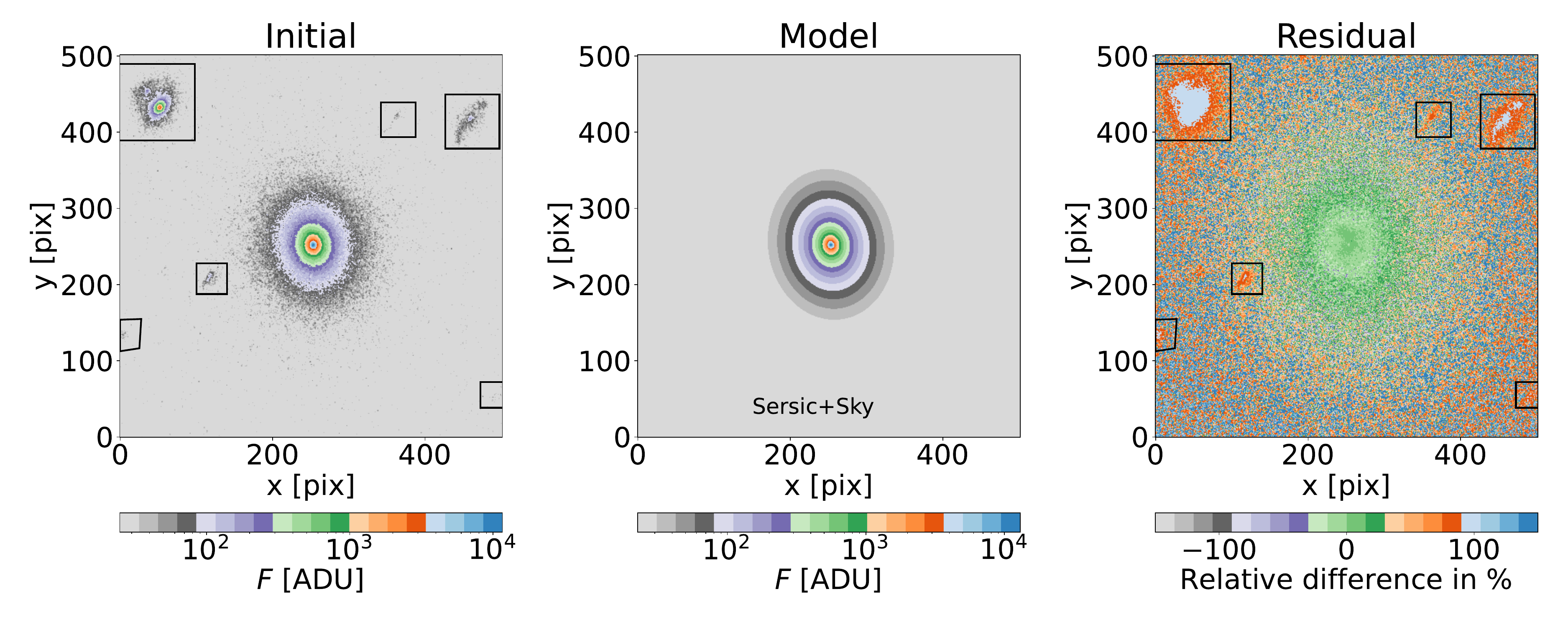}
\caption{
GALFIT analysis results for the host galaxy of radio galaxy 44. The model is S\'ersic+Sky, with best-fit S\'ersic parameters $n=4.84\pm0.01$ and $R_{e}=59.3\pm0.3\hspace{0.1cm}\mathrm{pix}$($\approx8.8\hspace{0.1cm}\mathrm{kpc}$). Image scales and markings are the same as in Fig.~\ref{Fig:10913_galfit}.}
\label{Fig:44_galfit}
\end{figure*}

In case of the host galaxy of radio galaxy 44 we model its light using a single S\'ersic profile. We also attempt to include a central PSF to account for the AGN, however, this two-component model fails to converge. The initial parameters of the S\'ersic model are estimated by inspecting the galaxy in DS9. We also include the Sky component. All additional optical sources in the image were masked. The S\'ersic+Sky model gives a good fit with the reduced $\chi^{2}$ value of $\sim0.78$. GALFIT finds the S\'ersic index value and the effective radius of the host galaxy to be $n=4.84\pm0.01$ and $R_{e}=59.3\pm0.3\hspace{0.1cm}\mathrm{pix}$ ($\approx 8.8\hspace{0.1cm}\mathrm{kpc}$, using spectroscopic redshift). The full list of the GALFIT output model parameters is in Table~\ref{table:galfit_output_44}, and the resulting GALFIT cube (initial, model, and residual image) is shown in Fig.~\ref{Fig:44_galfit}.
\begin{figure*}[!h]
\centering
\includegraphics[width=0.79\textwidth]{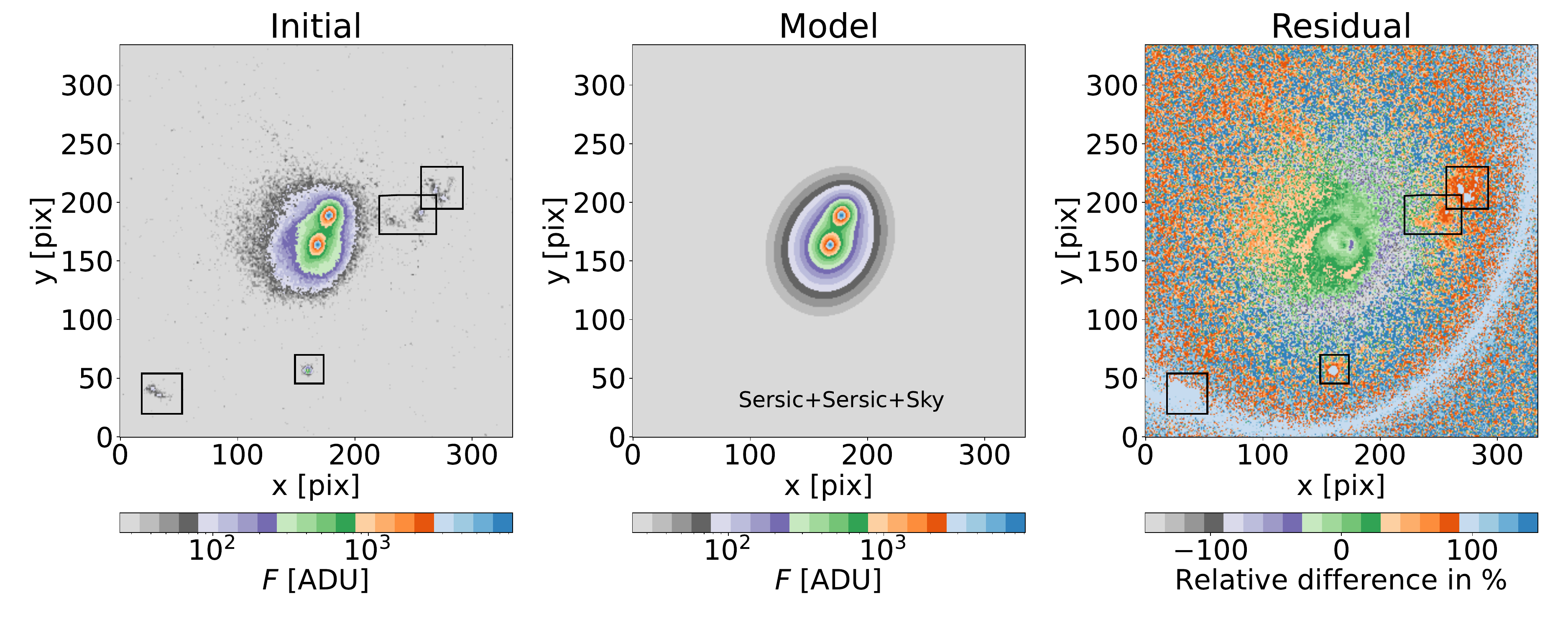}
\caption{GALFIT analysis results for the host galaxy of radio source 4092. The model is S\'ersic+S\'ersic (for another source) +Sky. The residual image reveals a spiral-like structure (see \ref{sec:galfit_results_4092}). Image scales and markings are the same as in Fig.~\ref{Fig:10913_galfit}.}
\label{Fig:4092_galfit}
\end{figure*}
\subsection{4092: Model, initial parameters, and results}
\label{sec:galfit_results_4092}
We attempt to model the host of radio source 4092 with a S\'ersic model. Given this source was not detected in X-ray (\ref{sec:xray_ps_within_the_group}), in this case, we do not include a PSF (AGN) component in the fit. We do, however, include another S\'ersic profile to account for the light from an additional nearby source, and we simultaneously fit the sky. The additional source is a galaxy listed in the \textit{ACS I-band Photometry Catalog} of optical sources \citep{leauthaud2007}. This source has no counterpart in the new catalog of spectroscopic redshifts \citep{Khostovan2025}, however, we find a high-confidence ($Q_f=4$) spectroscopic redshift (M. Salvato, priv. comm.) to be $\mathrm{spec\_z}=0.35289$. This value closely matches the spectroscopic redshift of the host of 4092 (Table \ref{table:srces}). All other sources in the cutout were masked. The reduced $\chi^{2}$ value of the fit is $\sim1.35$. The resulting GALFIT cube is shown in Fig.~\ref{Fig:4092_galfit} and the full list of the GALFIT output model parameters is in Table~\ref{table:galfit_output_4092}. The results of fitting point to the conclusion that this S\'ersic+S\'ersic+Sky model fails to describe the light distribution in this case. Inspecting the residual image in Fig.~\ref{Fig:4092_galfit}, we find that while the additional source is relatively well described
\begin{table} 
\caption{Best-fit (output) parameters for all components of S\'ersic+PSF+S\'ersic+Sky model used in GALFIT analysis of the host galaxy of WAT radio galaxy 10913.}              
\centering
\begin{tabular}{p{5cm} c} 
\hline\hline       
\multicolumn{2}{c}{Host galaxy ($1^{\mathrm{st}}$ S\'ersic component)}\\
    Parameter      & Value \\
    \hline                    
    Position (x,y) [pix] & \begin{tabular}{@{}c@{}}($345.308 \pm 0.005$\\ $341.562 \pm 0.004$)\end{tabular}\\
    Integrated magnitude [mag] & $-29.966 \pm 0.002$ \\
    Effective radius $R_{e}$ [pix] & $102.8 \pm 0.3$ \\
    S\'ersic index $n$ & $3.825 \pm 0.006$ \\
    Axis ratio ($b/a$) & $0.7274 \pm 0.0004$ \\
    Position angle [deg] & $-50.02 \pm 0.07$ \\ \hline \hline
\multicolumn{2}{c}{Central AGN (PSF component)}\\
    Parameter      & Value \\
    \hline                    
    Position (x,y) [pix] & \begin{tabular}{@{}c@{}}($339.58 \pm 0.02$,\\$342.52 \pm 0.02$)\end{tabular}\\
    Total magnitude [mag] & $-22.25 \pm 0.02$ \\ \hline \hline
\multicolumn{2}{c}{Additional source ($2^{\mathrm{nd}}$ S\'ersic component)}\\
    Parameter      & Value \\
    \hline                    
    Position (x,y) [pix] & \begin{tabular}{@{}c@{}} ($335.373\pm0.008$,\\$284.697\pm0.009$)\end{tabular} \\
    Integrated magnitude [mag] & $-26.211 \pm 0.003$ \\
    Effective radius $R_{e}$ [pix] & $7.56 \pm 0.03$ \\
    S\'ersic index $n$ & $1.759 \pm 0.009$ \\
    Axis ratio ($b/a$) & $0.801 \pm 0.002$ \\
    Position angle [deg] & $14.4 \pm 0.5$ \\ \hline \hline
\multicolumn{2}{c}{Sky background (Sky component)}\\
    Parameter      & Value \\
    \hline
    Sky background at center of fitting region [ADU] & $3.15 \pm 0.03$ \\
    Sky gradient in x [ADU/pix] & $(-1.5 \pm 0.1)\times 10^{-3}$ \\
    Sky gradient in y [ADU/pix] & $(-0.1 \pm 0.1)\times 10^{-3}$ \\
\hline
\end{tabular}
\label{table:galfit_output_10913}  
\end{table}
\begin{table} 
\caption{Best-fit (output) parameters for all components of S\'ersic+Sky model used in GALFIT analysis of the host galaxy of tailed radio galaxy 44.}
\centering
\begin{tabular}{p{5cm} c} 
\hline\hline       
\multicolumn{2}{c}{Host galaxy (S\'ersic component)}\\
    Parameter      & Value \\
    \hline                    
    Position (x,y) [pix] & \begin{tabular}{@{}c@{}}($253.974 \pm 0.004$\\ $252.488 \pm 0.005$)\end{tabular}\\
    Integrated magnitude [mag] & $-29.006 \pm 0.003$ \\
    Effective radius $R_{e}$ [pix] & $58.9 \pm 0.3$ \\
    S\'ersic index $n$ & $4.82 \pm 0.01$ \\
    Axis ratio ($b/a$) & $0.8303 \pm 0.0008$ \\
    Position angle [deg] & $8.8 \pm 0.2$ \\ \hline \hline

\multicolumn{2}{c}{Sky background (Sky component)}\\
    Parameter      & Value \\
    \hline
    Sky background at center of fitting region [ADU] & $2.28 \pm 0.03$ \\
    Sky gradient in x [ADU/pix] & $(3 \pm 2)\times 10^{-4}$ \\
    Sky gradient in y [ADU/pix] & $(3.8 \pm 0.1)\times 10^{-3}$ \\
\hline
\end{tabular}
\label{table:galfit_output_44}  
\end{table}
with a single S\'ersic model, there are bright structures present around the host galaxy of radio source 4092, which cannot be fitted away with a S\'ersic profile. We suspect these may be either the spiral arms or a ring-like structure. Given our galaxy has a companion (the neighboring galaxy described above) the first case seems to be more likely. Namely, the gravitational interaction with the close companion can drive density waves in the galaxy causing orbits' distortion and resulting in 2 arms spiral structure \citep{LS1964, KN1979}. Moreover, asymmetry is noticeable in the residual image with the excess of light on one edge of the galaxy and scarcity on the opposite side. We believe this asymmetry is also related to gravitational interaction with the close companion, that is, is a consequence of a potential galaxy merger.

\vspace{-0.1cm}
\subsection{GALFIT output parameters: full lists}
\label{sec:full_lists_params}
The results of fitting different models to the galaxy light distribution data for hosts of the three radio sources are presented in the above sections. Here we provide a full list of parameter values resulting from different fits in Tables~\ref{table:galfit_output_10913}, \ref{table:galfit_output_44}, and \ref{table:galfit_output_4092} for hosts of 10913, 44, and 4092, respectively. The fitting was performed using GALFIT in the pixel space, and the resulting positions and lengths are presented in pixels. The plate scale of the data used is $0.03\hspace{0.1cm}^{\prime\prime}\mathrm{/pixel}$ and the image was convolved with the $0.095^{\prime\prime}$ full width at half maximum (FWHM) PSF image prior to fitting. Integrated magnitude is as defined in Equation (5) of GALFIT User's manual\footnote{\url{https://users.obs.carnegiescience.edu/peng/work/galfit/README.pdf}}, where ST magnitude system zero point (PHOTZPT=-21.1) was used.

\vspace{-2cm}
\begin{table}  
\caption{Best-fit (output) parameters for all components of S\'ersic+S\'ersic+Sky model used in GALFIT analysis of the host galaxy of radio source 4092.} 
\centering
\begin{tabular}{p{5cm} c} 
\hline\hline       
\multicolumn{2}{c}{Host galaxy ($1^{\mathrm{st}}$ S\'ersic component)}\\
    Parameter      & Value \\
    \hline                    
    Position (x,y) [pix] & \begin{tabular}{@{}c@{}}($169.369 \pm 0.008$\\ $164.334 \pm 0.008$)\end{tabular}\\
    Integrated magnitude [mag] & $-28.170 \pm 0.008$ \\
    Effective radius $R_{e}$ [pix] & $49.0 \pm 0.6$ \\
    S\'ersic index $n$ & $4.62 \pm 0.03$ \\
    Axis ratio ($b/a$) & $0.838 \pm 0.002$ \\
    Position angle [deg] & $-23.7 \pm 0.4$ \\ \hline \hline
\multicolumn{2}{c}{Additional source ($2^{\mathrm{nd}}$ S\'ersic component)}\\
    Parameter      & Value \\
    \hline                    
    Position (x,y) [pix] & \begin{tabular}{@{}c@{}} ($179.014 \pm 0.006$,\\$189.512 \pm 0.007$)\end{tabular} \\
    Integrated magnitude [mag] & $-26.576 \pm 0.003$ \\
    Effective radius $R_{e}$ [pix] & $6.95 \pm 0.03$ \\
    S\'ersic index $n$ & $2.52 \pm 0.01$ \\
    Axis ratio ($b/a$) & $0.806 \pm 0.003$ \\
    Position angle [deg] & $-32.2 \pm 0.5$ \\ \hline \hline
\multicolumn{2}{c}{Sky background (Sky component)}\\
    Parameter      & Value \\
    \hline
    Sky background at center of fitting region [ADU] & $-1.69 \pm 0.06$ \\
    Sky gradient in x [ADU/pix] & $(-1.13 \pm 0.03)\times 10^{-2}$ \\
    Sky gradient in y [ADU/pix] & $(1.58 \pm 0.03)\times 10^{-2}$ \\
\hline
\end{tabular}
\label{table:galfit_output_4092} 
\end{table}
\clearpage

\section{SBPs of hosts of radio galaxies 10913 and 44: Elliptical isophote fits}
\label{app:sbps}
To generate SBPs, elliptical isophotes were fitted to the HST-ACS F814W images \citep{koekemoer2007} of the hosts of radio galaxies 10913 and 44 using the Python-based \texttt{Photutils} package \citep{bradley25}. In both cases, sources other than the galaxy itself were masked prior to fitting, and the fitting was restricted to avoid the central PSF-size region (assuming PSF FWHM=$0.095^{\prime\prime}$). The latter was done to exclude any potential AGN flux from the fit. As input, the algorithm requires geometric parameters of an initial ellipse, used to begin the numerical fitting procedure to find isophotes at different semi-major axis lengths $a$. It is important the provided center of the initial ellipse is highly accurate since the algorithm itself has no way to determine a galaxy's center. Therefore, for the central coordinates we use the output of GALFIT analysis (see Tables~\ref{table:galfit_output_10913} and \ref{table:galfit_output_44} in appendix \ref{app:Galfit}, cutout sizes used here are the same as in GALFIT analysis). By closely inspecting the galaxy cutout in DS9 \citep{DS9} we find the other geometric parameters for the initial ellipse ($a$, ellipticity, and the angle). \citet{busko1996} showed that for the method of \citet{jedrzejewski1987}, the precision of isophote fitting depends on the local radial gradient relative error. They find that gradient errors larger than $50\%$ significantly decrease the accuracy of fit and can result in underestimated errors for isophote geometry parameters. Here, we conservatively set the maximum acceptable relative error in the local radial intensity gradient to $10\%$ (maxgerr=0.1). We choose the geometric growing mode for the $a$ parameter with the step value 0.08. The algorithm works in pixel space and since the initial cutouts are provided in units of electron counts it calculates the surface brightness (corresponding to the mean intensity along an elliptical isophote) in electrons/pixel for each isophote, taking into account, namely, correcting for the area of masked regions (for details see \citealt{jedrzejewski1987}). We convert the values of surface brightness to $\mathrm{mag}/\mathrm{arcsec^2}$, where magnitudes are in ST system, using information on sensitivity, photometric zero point and plate scale of the imaging data. We also calculate the corresponding surface brightness errors in $\mathrm{mag}/\mathrm{arcsec^2}$ from the errors originating directly from the isophote fit via error propagation. The commonly used analytical form of S\'ersic profile is given in terms of equivalent radius $R$. Therefore, we create equivalent circular profile by converting the semi-major axis lengths of fitted isophotes to radius using $R=a\cdot \sqrt{1-e}$, where $e$ is isophote's ellipticity, and we convert the units from pixels to kpc, using the galaxy's spectroscopic redshift (see Table~\ref{table:srces}).
\section{Rotation angles and projection of radio galaxy 10913}
\label{app:rotation}
\begin{figure*}[h]
\centering
\includegraphics[width=0.77\textwidth]{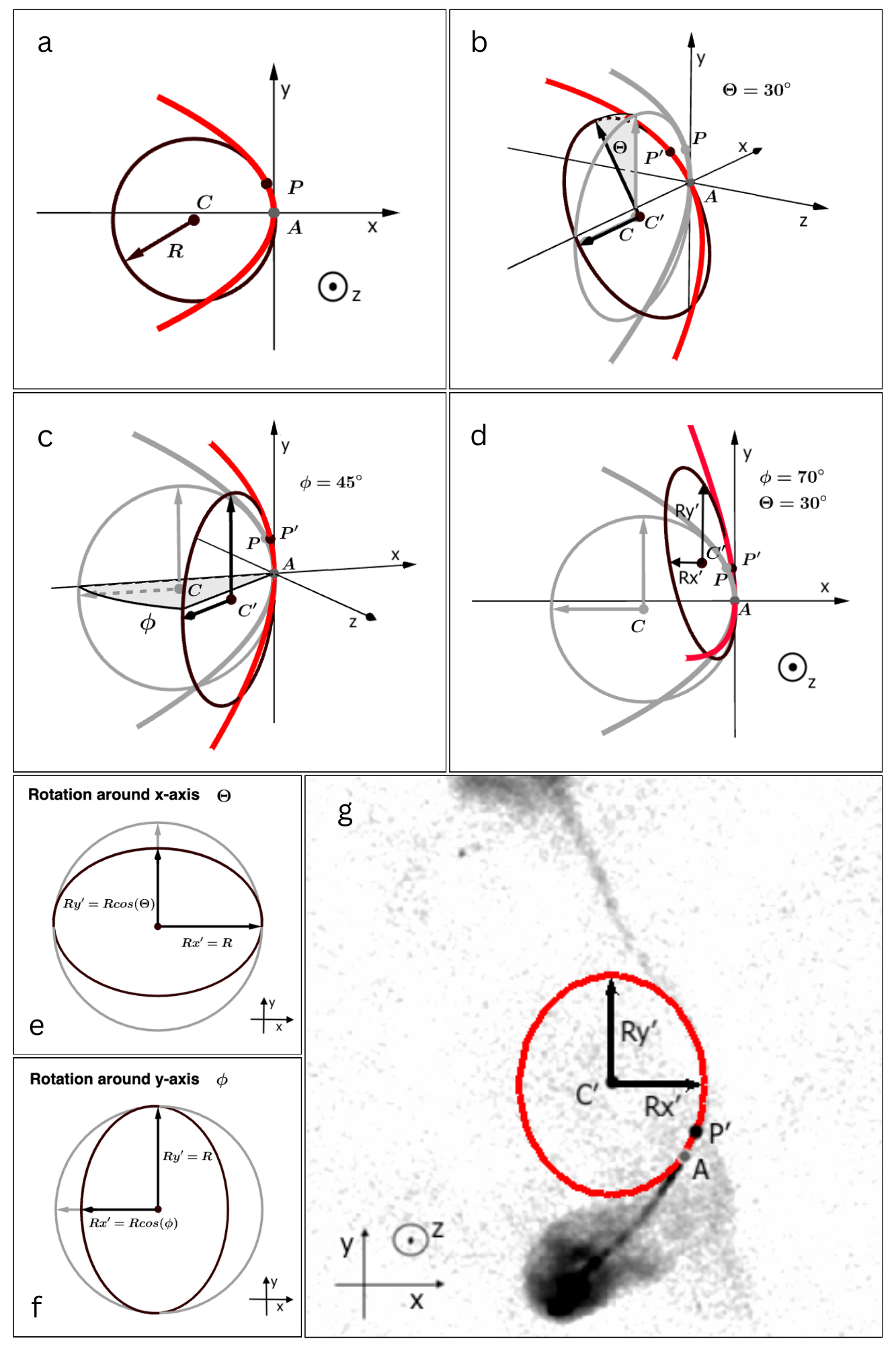}
\caption{Illustration of projection effects due to WAT orientation (rotation). In panels a - f the geometry of the problem is visualized and explained step by step (follow the text of Appendix \ref{app:rotation}), and in panel g the projection is analyzed for the real case - radio galaxy 10913. In all panels x-y plane corresponds to the plane of the sky, and z axis aligns with the LoS (pointing towards the observer). Markings $A$, $P$, $C$ and $R$ denote the radio core, the bending point, the center and the radius of the osculating circle at the bending point, respectively. $P^{\prime}$ and $C^{\prime}$ are the projected locations of $P$ and $C$ after rotation. Panel a:  WAT lies entirely in the plane of the sky, $R$ - radius of the osculating circle at the bending point, $P$, can be extracted directly from the observations; Panel b: WAT rotated by angle $\Theta$ around x-axis (tilted from the plane of the sky); gray indicates original jet position and osculating circle; Panel c: WAT rotated by angle $\phi$ around y-axis; Panel d: WAT rotated by $\Theta$ (x) and $\phi$ (y), projected osculating circle transformed into an ellipse; Panels e and f: Effect of $\Theta$ (e) and $\phi$ (f) rotations on x and y components of the projected $R$; Panel g: Observed radio morphology of WAT 10913, measuring projected x and y components of $R$.}
\label{fig:rotation_problem}
\end{figure*}

We analyze the rotation angles of WAT radio galaxy 10913 to estimate the projection effects and correct for them in the analysis of radio jet bending (\ref{sec:bending_10913}). The geometry of the rotation problem is shown in Fig.~\ref{fig:rotation_problem}, where the plane of the sky is spanned by x and y axes, and z axis points towards the observer, namely, coincides with the LoS.\par
We first consider an ideal case, where WAT radio structure lies entirely in the plane of the sky (x-y plane) as shown in panel a of Fig.~\ref{fig:rotation_problem}. In this case, no projection effects take place and the geometry of the jets, namely, the radius of curvature, $R$, (radius of the osculating circle at the bending point) can be directly extracted from observations.\par
There are three angles of rotation, each around one of the three axes, which can affect how we see the radio structure. Only two of these three rotations introduce the LoS component, that is, projection effects, here defined as: $\Theta$ - measuring the rotation of the radio structure around x-axis in the counterclockwise direction starting from the x-y plane (visualized in panel b of Fig.~\ref{fig:rotation_problem}), and $\phi$ - rotation around y-axis in the counterclockwise direction starting from the x-y plane (visualized in panel c of Fig.~\ref{fig:rotation_problem}). The third angle of rotation is around z axis in the counterclockwise direction starting from the x-z plane, that is, this rotation happens entirely in the plane of the sky. From radio image, we find this angle is approximately $0$.\par
Both rotations ($\Theta$ and $\phi$) change the plane of the sky projection of the radio structure, namely, $R$, thereby only preserving the length of components which lie along the corresponding rotation axes as shown in panels e (for $\Theta$) and f (for $\phi$) in Fig.~\ref{fig:rotation_problem}. The rotations combined transform the (non-projected) osculating circle fitting the curvature at the bending point into a new projected shape:\ an ellipse (see panel d in Fig.~\ref{fig:rotation_problem}). Extracting measures of this ellipse along x and y axes, and combining them with basic $\Theta$ angle constraints originating directly from observations, allows us to place constraints on the other angle $\phi$ and non-projected value of $R$.\par
In this particular case, namely, for WAT radio galaxy 10913 we measure the projected values of radius $R$ at the bending point along x and y axes: $R_x^{\prime}\sim65\hspace{0.1cm}\mathrm{kpc}$ and $R_y^{\prime}\sim77\hspace{0.1cm}\mathrm{kpc}$ as shown in panel g of Fig.~\ref{fig:rotation_problem} (red ellipse). The projected (observed) location of the bending point (labeled as $P^{\prime}$ in panel g) is set to correspond to a hotspot found along the northern jet very close to the central core (labeled as $A$). From the geometry of the problem as described above:
\begin{equation}
\label{eq:ry_prime_rotation}
    R=\frac{R_y^{\prime}}{cos\Theta}
\end{equation}
For this particular radio galaxy we know $\Theta\in[0^{\circ},45^{\circ}]$ because we observe both jets and lobes clear, and the bottom jet and lobe seem to be experiencing relativistic boosting. This yields $R\in[\sim77,\sim109]\hspace{0.1cm}\mathrm{kpc}$. In the x direction,
\begin{equation}
    R=\frac{R_x^{\prime}}{cos\phi}
\end{equation}
This, combined with the above expression \ref{eq:ry_prime_rotation}, gives us a constraint on the angle $\phi,$
\begin{equation}
    cos\phi=\frac{R_x^{\prime}}{R_y^{\prime}}cos\Theta
.\end{equation}
Using the measured projected values of $R$ and $\Theta\in[0^{\circ},45^{\circ}]$, we find $\phi\in[\sim32^{\circ}, \sim 53^{\circ}]$, where the first value corresponds to the case when $\Theta=0^{\circ}$ and for the second one $\Theta=45^{\circ}$. No matter the symmetry of the $cos$ function, we reject negative solutions for $\phi$ because, in combination with the allowed $\Theta$ range, they cannot produce the observed radio morphology, but rather the opposite one (symmetric with respect to x-axis). These constraints on the value of $R$, $\Theta$ and $\phi$ are then used in the jet bending analysis and calculation of the galaxy's peculiar velocity, and its velocity with respect to the IGM as described in Section \ref{sec:bending_10913}.
\end{appendix}

\end{document}